\documentclass[journal]{IEEEtran}

\usepackage{cite}
\usepackage{amsmath,amssymb,amsfonts}
\usepackage{textcomp}

\usepackage{graphicx}
\usepackage{subfig}
\usepackage{epstopdf}
\usepackage{comment}
\usepackage{amsthm}
 
\newtheoremstyle{mystl}
  {0}   
  {0}   
  {\normalfont}  
  {\parindent}       
  {\itshape} 
  {:}         
  {5pt plus 1pt minus 1pt} 
  {\thmname{#1} \thmnumber{#2}}          

\theoremstyle{mystl}
\newtheorem{rmk}{Remark}
\newtheorem{lem}{Lemma}
\newtheorem{thme}{Theorem}
\newtheorem{dfn}{Definition}
\newtheorem{prop}{Proposition}
\newtheorem{ass}{Assumption}
\newtheorem{cllry}{Corollary}

\def\BibTeX{{\rm B\kern-.05em{\sc i\kern-.025em b}\kern-.08em
    T\kern-.1667em\lower.7ex\hbox{E}\kern-.125emX}}
\begin{document}
\title{Observer-based Event-triggered Boundary Control of a Class of Reaction-Diffusion PDEs}
\author{Bhathiya Rathnayake, Mamadou Diagne, Nicol\'as Espitia, and Iasson Karafyllis
\thanks{B. Rathnayake is with the Department of Electrical, Computer, and Systems Engineering, Rensselaer Polytechnic Institute, New York, 12180, USA. Email: rathnb@rpi.edu}
\thanks{M. Diagne is with the Department of Mechanical, Aerospace, and Nuclear Engineering, Rensselaer Polytechnic Institute, New York, 12180, USA. Email: diagnm@rpi.edu}
\thanks{N. Espitia is with CRIStAL UMR 9189 CNRS - Centre
de Recherche en Informatique Signal et Automatique de Lille - CNRS, Centrale Lille, Univ. Lille, F-59000
Lille, France. Email: nicolas.espitia-hoyos@univ-lille.fr}
\thanks{I. Karafyllis is with the Department of Mathematics, National Technical University of Athens, Greece. Email: iasonkar@central.ntua.gr}}
\maketitle

\begin{abstract}
This paper presents an observer-based event-triggered boundary control strategy for a class of reaction-diffusion PDEs with Robin actuation. The observer only requires boundary measurements. The control approach consists of a backstepping output feedback boundary controller, derived using estimated states, and a dynamic triggering condition, which determines the time instants at which the control input needs to be updated. It is shown that under the proposed observer-based event-triggered boundary control approach, there is a minimal dwell-time between two triggering instants independent of initial conditions. Furthermore, the well-posedness and the global exponential convergence of the closed-loop system to the equilibrium point are established. A simulation example is provided to validate the theoretical developments.
\end{abstract}


\section{Introduction}
Event-triggered control (ETC) is a control implementation technique that closes the feedback loop only if an event indicates that the control input error exceeds an appropriate threshold. When an event occurs, the controller computes the control value and transmits it to the actuator, completing the feedback path. Therefore, unlike periodic sampled-data control \cite{chen1991input,aastrom2013computer}, ETC only requires the control input to be updated aperiodically (only when needed). For networked and embedded control systems, the periodic computation and transmission of the control inputs are sometimes not desirable due to task scheduling limitations and bandwidth constraints in the communication \cite{walsh2001scheduling,wang2010event,peng2013event,nowzari2019event}. Despite the developments in aperiodic sampled-data control \cite{hetel2017recent,karafyllis2017sampled,karafyllis2018sampled,kang2018distributed}, they lack explicit criteria for selecting appropriate sampling schedules. ETC, on the other hand,  provides a rigorous \textit{resource-aware} method of implementing the control laws into digital platforms \cite{heemels2012introduction,yao2013resource,lemmon2010event}.

In general, ETC consists of two main components: a feedback control law that stabilizes the system and an event-triggered mechanism, which determines when the control value has to be computed and sent toward the actuator. In the literature, one can find two main approaches to the control law design: \textit{emulation} (e.g. \cite{heemels2012introduction}) and \textit{co-design} (e.g. \cite{peng2013event}). The former requires a pre-designed continuous feedback controller applied to the plant in a Zero-Order-Hold fashion between two event times. In co-design,  the feedback control law and the event-triggered mechanism are simultaneously designed to obtain the desired stability properties. An important property that every ETC design should possess is the non-existence of \textit{Zeno behavior}\cite{lemmon2010event}; otherwise, it will lead to the triggering of an infinite number of control updates over a finite period, making the design infeasible for digital implementation. Usually, ETC designs are ensured to be Zeno-free by showing the existence of a guaranteed lower bound for the time between two consecutive events, known as the minimal dwell-time.

In the case of finite-dimensional systems, ETC has grown to be a mature field of research. During the past  decade, many significant results related to ETC have been reported on systems described by linear and nonlinear  ODEs in both full-state and output feedback settings (see \cite{tabuada2007event,heemels2012introduction,6069816,heemels2012periodic,marchand2012general,tallapragada2013event,postoyan2014framework,girard2014dynamic,abdelrahim2015stabilization,borgers2018periodic,wang2019periodic,zhang2020event,zhang2020systematic}). Recently, there has been a growing interest in periodic ETC \cite{heemels2012periodic,borgers2018periodic,wang2019periodic}, which employs a sampled-data event-triggered mechanism instead of the usual continuous-time event trigger. With this, the system states and the triggering condition need to be monitored and evaluated only at the sampling instants, making ETC more realistic for digital implementation. Despite some recent developments such as in \cite{selivanov2016distributed,espitia2016event,espitia2017stabilization,espitia2017event,wang2019observer,espitia2020observer,espitia2020event,diagne2020event}, ETC strategies for PDE systems have not reached the level of maturity seen by finite-dimensional systems yet. Only recently, even the notions of solutions of linear hyperbolic and parabolic PDEs under sampled-data control have been clarified \cite{karafyllis2017sampled,karafyllis2018sampled}. 

One can find several recent works that employ event-triggered boundary control strategies based on emulation on hyperbolic and parabolic PDEs \cite{espitia2016event,espitia2017event,espitia2020observer,espitia2020event}. The authors of \cite{espitia2016event} propose an output feedback event-triggered boundary controller for 1-D  linear hyperbolic systems of conservation laws, using Lyapunov techniques. By utilizing a dynamic triggering condition,  an event-based backstepping boundary controller is designed for a coupled $2\times 2$ hyperbolic system in \cite{espitia2017event}. This work is extended in \cite{espitia2020observer} to obtain an event-triggered output feedback boundary controller for a similar system. In the case of parabolic PDEs, \cite{espitia2020event} is the first work that reports an event-based boundary control design. Using ISS properties and small gain arguments, the authors propose a full state feedback backstepping ETC strategy for a reaction-diffusion system with constant parameters and Dirichlet boundary conditions. 

The importance of event-triggered output feedback control cannot be stressed enough as the use of full state measurements is either impossible or prohibitively expensive for many practical applications. Some studies such as \cite{selivanov2016distributed,xue2016output,ge2020observer} report several event-triggered output feedback designs for parabolic PDEs. All these works, however, rely on in-domain control and distributed observation. Event-based boundary control of parabolic PDEs with boundary sensing only is quite challenging, and no prior results have been reported. The possibility of avoiding Zeno behavior in ETC of general parabolic PDE systems with boundary actuation and boundary observation is not known. The actuation type is critical as both Dirichlet and Neumann actuation pose a severe impediment in establishing a minimal dwell-time and hence well-posedness and convergence results, due to unbounded local terms. However, we have identified that a class of reaction-diffusion equations with Robin actuation is conducive for event-triggered boundary control with boundary sensing. This paper proposes an observer-based event-triggered backstepping boundary controller for the class of PDEs mentioned above using emulation. The observer only requires boundary observation. The main contributions are as \vspace{-2pt}follows:
\begin{itemize}
\item We consider a class of reaction-diffusion systems with Robin actuation. We perform emulation on an observer-based backstepping boundary control design and propose a dynamic triggering condition under which Zeno behavior is excluded. It is proved the existence of a minimal-dwell-time independent of the initial conditions.
\item We prove the well-posedness of the closed-loop system and its global exponential convergence to the equilibrium point in $L^{2}$-sense.\end{itemize}

The paper is organized as follows. Section 2 introduces the class of linear reaction-diffusion system and the continuous-time output feedback boundary control. Section 3 presents the observer-based event-triggered boundary control and some properties. In Section 4, we discuss the main results of this paper. We provide a numerical example in Section 5 to illustrate the results and conclude the paper in Section 6.

\textit{Notation:} $\mathbb{R}_{+}$ is the nonnegative real line whereas $\mathbb{N}$ is the set of natural numbers including zero.  By $C^{0}(A;\Omega)$, we denote the class of continuous functions on $A\subseteq\mathbb{R}^{n}$, which takes values in $\Omega\subseteq\mathbb{R}$. By $C^{k}(A;\Omega)$, where $k\geq 1$, we denote the class of continuous functions on $A$, which takes values in $\Omega$ and has continuous derivatives of order $k$.  $L^{2}(0,1)$ denotes the equivalence class of Lebesgue measurable functions $f:[0,1]\rightarrow\mathbb{R}$ such that $\Vert f\Vert=\big(\int_{0}^{1}\vert f(x)\vert^{2}\big)^{1/2}<\infty$. Let $u:[0,1]\times\mathbb{R}_{+}\rightarrow\mathbb{R}$ be given. $u[t]$ denotes the profile of $u$ at certain $t\geq 0$, \textit{i.e.,} $\big(u[t]\big)(x)=u(x,t),$ for all $x\in [0,1]$. For an interval $I\subseteq\mathbb{R}_{+},$ the space $C^{0}\big(I;L^{2}(0,1)\big)$ is the space of continuous mappings $I\ni t\rightarrow u[t]\in L^{2}(0,1)$. $I_{m}(\cdot), $ and $J_{m} (\cdot)$ with $m$ being an integer respectively denote the modified Bessel and (nonmodified) Bessel functions of the first kind.   
\section{ Observer-Based Backstepping Boundary Control and Emulation}

Let us consider the following 1-D reaction-diffusion system with constant coefficients:
\begin{subequations}\label{ctp}
\begin{align}\label{ctpe1}
u_{t}(x,t)&=\varepsilon u_{xx}(x,t)+\lambda u(x,t),\\
\label{ctpe2}
u_{x}(0,t)&=0,\\\label{ctpe3}
u_{x}(1,t)+qu(1,t)&=U(t),
\end{align}
\end{subequations}
and the initial condition $u[0]\in L^{2}(0,1),$ where $\varepsilon,\lambda>0,$ $u: [0,1]\times [0,\infty)\rightarrow\mathbb{R}$ is the system state and $U(t)$ is the control input.\smallskip
\begin{ass}\label{ass1}
$q>(\lambda+\varepsilon)/2\varepsilon.$\smallskip
\end{ass}

\begin{rmk}\label{rem1}
It should be mentioned that the solution of the PDE system defined as \eqref{ctp} with $U(t)=0$ (zero input), where $\varepsilon, q>0$ and $\lambda\in\mathbb{R}$ are constants, satisfies the estimate
\begin{equation}\label{est}
\Vert u[t]\Vert\leq e^{(\lambda-\varepsilon\mu^{2})t}\Vert u[0]\Vert,
\end{equation}
for all $t\geq 0$ and $u[0]\in L^{2}(0,1),$ where $\mu$ is the unique solution of the transcendental equation $\mu\tan(\mu)=q$ in the interval $(0,\pi/2)$ \cite{polyanin2015handbook}. The estimate \eqref{est} guarantees the global exponential stability of the zero-input system in $L^{2}$-norm when $\lambda<\varepsilon\mu^2$. Further, an eigenfunction expansion of its solution shows that the system is unstable when $\lambda>\varepsilon\mu^2$. Since $\mu$ is in the interval $(0,\pi/2)$, this implies instability when $\lambda>\varepsilon\pi^2/4$, no matter what $q>0$ is.\smallskip
\end{rmk}

We propose an observer for the system \eqref{ctp} using $u(0,t)$ as the available measurement/output. Note that the output is anticollocated with the input. The observer consists of a copy of the system \eqref{ctp} with output injection terms, which is stated as follows:
\begin{subequations}\label{cto}
\begin{align}\label{ctoe1}
\begin{split}
\hat{u}_{t}(x,t)&=\varepsilon \hat{u}_{xx}(x,t)+\lambda \hat{u}(x,t)\\&\qquad+p_1(x)\big(u(0,t)-\hat{u}(0,t)\big),
\end{split}
\\\label{ctoe2}
\hat{u}_{x}(0,t)&=p_{10}\big(u(0,t)-\hat{u}(0,t)\big),
\\\label{ctoe3}
\hat{u}_{x}(1,t)+q\hat{u}(1,t)&=U(t),
\end{align}
\end{subequations}
and the initial condition $\hat{u}[0]\in L^{2}(0,1)$. Here, the function $p_{1}(x)$ and the constant $p_{10}$ are observer gains to be determined. Let us denote the observer error by $\tilde{u}(x,t)$, which is defined as\begin{equation}\label{diff}
\tilde{u}(x,t):=u(x,t)-\hat{u}(x,t).
\end{equation}
By subtracting \eqref{cto} from \eqref{ctp}, one can see that $\tilde{u}(x,t)$ satisfies the following PDE:
\begin{subequations}\label{ctoe}
\begin{align}\label{ctoee1}
\begin{split}
\tilde{u}_{t}(x,t)&=\varepsilon\tilde{u}_{xx}(x,t)+\lambda\tilde{u}(x,t)-p_{1}(x)\tilde{u}(0,t),
\end{split}
\\\label{ctoee2}
\tilde{u}_{x}(0,t)&=-p_{10}\tilde{u}(0,t),
\\\label{ctoee3}
\tilde{u}_{x}(1,t)+q\tilde{u}(1,t)&=0.
\end{align}
\end{subequations}
\begin{prop}\label{prop1} 
Subject to Assumption \ref{ass1} and the invertible backstepping transformation\begin{equation}\label{ctobt}
\tilde{u}(x,t)=\tilde{w}(x,t)-\int_{0}^{x}P(x,y)\tilde{w}(y,t)dy,
\end{equation}where \begin{equation}\label{solP}
\begin{split}
P(x,y)=&\frac{q\lambda/\varepsilon}{\sqrt{\lambda/\varepsilon+q^{2}}}\\&\times\int_{0}^{x-y}e^{-q\tau/2}I_{0}\Big(\sqrt{\lambda(2-x-y)(x-y-\tau)/\varepsilon}\Big)\\&\hspace{35pt}\times\sinh\Big(\frac{\sqrt{\lambda/\varepsilon+q^{2}}}{2}\tau\Big)d\tau\\&
-\frac{\lambda}{\varepsilon}(1-y)\frac{I_{1}\Big(\sqrt{\lambda\big((1-y)^{2}-(1-x)^{2}\big)/\varepsilon}\Big)}{\sqrt{\lambda\big((1-y)^{2}-(1-x)^{2}\big)/\varepsilon}},
\end{split}
\end{equation}for $0\leq y\leq x\leq 1$, then, the system \eqref{ctoe} with the gains $p_{1}(x)$ and $p_{10}$ chosen as\begin{align}\label{p1}
p_{1}(x)=\varepsilon P_{y}(x,0),\hspace{20pt}p_{10}=P(0,0)=-\frac{\lambda}{2\varepsilon},
\end{align}
gets transformed to the following globally $L^{2}$-exponentially stable observer error target system 
\begin{subequations}\label{ctots}
\begin{align}\label{ctotse1}
\tilde{w}_{t}(x,t)&=\varepsilon \tilde{w}_{xx}(x,t),
\\\label{ctotse2}
\tilde{w}_{x}(0,t)&=0,
\\\label{ctotse3}
\tilde{w}_{x}(1,t)&=-q\tilde{w}(1,t).
\end{align}
\end{subequations}
\end{prop}

\textit{Proof:} See Appendix A.\smallskip

The inverse transformation of \eqref{ctobt} can be shown to be as follows:
\begin{equation}\label{ibtoe}
\tilde{w}(x,t)=\tilde{u}(x,t)+\int_{0}^{x}Q(x,y)\tilde{u}(y,t)dy,
\end{equation}
where $Q(x,y)$ is 
\begin{equation}\label{Qsol}
\begin{split}
Q(x,y)=&\frac{q\lambda/\varepsilon}{\sqrt{-\lambda/\varepsilon+q^{2}}}\\&\times\int_{0}^{x-y}e^{-q\tau/2}J_{0}\Big(\sqrt{\lambda(2-x-y)(x-y-\tau)/\varepsilon}\Big)\\&\hspace{35pt}\times\sinh\Big(\frac{\sqrt{-\lambda/\varepsilon+q^{2}}}{2}\tau\Big)d\tau\\&
-\frac{\lambda}{\varepsilon}(1-y)\frac{J_{1}\Big(\sqrt{\lambda\big((1-y)^{2}-(1-x)^{2}\big)/\varepsilon}\Big)}{\sqrt{\lambda\big((1-y)^{2}-(1-x)^{2}\big)/\varepsilon}},
\end{split}
\end{equation}
for $0\leq y\leq x\leq 1$.\smallskip

\begin{prop}\label{prop2} The invertible backstepping transformation
\begin{equation}\label{ctbt}
\hat{w}(x,t)=\hat{u}(x,t)-\int_{0}^{x}K(x,y)\hat{u}(y,t)dy,
\end{equation}
where\begin{equation}\label{ctcks}
K(x,y)=-\frac{\lambda}{\varepsilon}x\frac{I_{1}\big(\sqrt{\lambda(x^{2}-y^{2})/\varepsilon}\big)}{\sqrt{\lambda(x^{2}-y^{2})/\varepsilon}},
\end{equation} for $0\leq y\leq x\leq 1,$  and a control law $U(t)$ chosen as\begin{equation}\label{ctcl}
U(t)=\int_{0}^{1}\Big(rK(1,y)+K_{x}(1,y)\Big)\hat{u}(y,t)dy,
\end{equation}map the system \eqref{cto} with the gains $p_{1}(x)$ and $p_{10}$ chosen as in \eqref{p1}, into the following target system:
\begin{subequations}\label{etots}\begin{align}\label{etotse1}
\hat{w}_{t}(x,t)&=\varepsilon \hat{w}_{xx}(x,t)+g(x)\tilde{w}(0,t),
\\\label{etotse2}
\hat{w}_{x}(0,t)&=-\frac{\lambda}{2\varepsilon}\tilde{w}(0,t),
\\\label{etotse3}
\hat{w}_{x}(1,t)&=-r\hat{w}(1,t),
\end{align}\end{subequations}
with\begin{equation}\label{gt}
g(x)=p_{1}(x)-\frac{\lambda}{2}K(x,0)-\int_{0}^{x}K(x,y)p_{1}(y)dy,
\end{equation}and
\begin{equation}\label{rt}
r=q-\frac{\lambda}{2\varepsilon}.
\end{equation}\end{prop}

\textit{Proof:} See Appendix B.\smallskip

The inverse transformation of \eqref{ctbt} can be shown to be as follows:\begin{equation}\label{ink}
\hat{u}(x,t)=\hat{w}(x,t)+\int_{0}^{x}L(x,y)\hat{w}(y,t)dy,
\end{equation}where
\begin{equation}\label{Lsol}
L(x,y)=-\frac{\lambda}{\varepsilon}x\frac{J_{1}\big(\sqrt{\lambda(x^{2}-y^{2})/\varepsilon}\big)}{\sqrt{\lambda(x^{2}-y^{2})/\varepsilon}},
\end{equation}
for $0\leq y\leq x\leq 1$.\smallskip

\begin{prop}\label{prop3}
Subject to Assumption \ref{ass1}, the closed-loop system which consists of the plant \eqref{ctp}  and the observer \eqref{cto} with the continuous-time control law \eqref{ctcl}, is globally exponentially stable in $L^{2}$-sense.
\end{prop}
\textit{Proof:} See Appendix C.

\subsection{Emulation of the Observer-based Backstepping Boundary Control}

We strive to stabilize the closed-loop system containing the plant \eqref{ctp} and the observer \eqref{cto} while sampling the continuous-time controller $U(t)$ given by \eqref{ctcl} at a certain sequence of time instants $(t_{j})_{j\in\mathbb{N}}$. These time instants will be given a precise characterization later on based on an event trigger. The control input is held constant between two successive time instants and is updated when a certain condition is met. Therefore, we define the control input for $t\in[t_{j},t_{j+1}),j\in\mathbb{N}$ as
\begin{equation}\label{etcl}
U_{j}:=U(t_{j})=\int_{0}^{1}\Big(rK(1,y)+K_{x}(1,y)\Big)\hat{u}(y,t_{j})dy.
\end{equation}Accordingly, the boundary conditions \eqref{ctpe3} and \eqref{ctoe3} are modified, respectively, as follows:\begin{equation}\label{mctpe3}
u_{x}(1,t)+qu(1,t)=U_{j},
\end{equation}\begin{equation}\label{mctoe3}
\hat{u}_{x}(1,t)+q\hat{u}(1,t)=U_{j}.
\end{equation}
The deviation between the continuous-time control law and its sampled counterpart, referred to as the input holding error, is defined as follows:\begin{equation}\label{dt}
\begin{split}
d(t):=\int_{0}^{1}\Big(rK(1,y)+K_{x}(1,y)\Big)\big(\hat{u}(y,t_{j})-\hat{u}(y,t)\big)dy.
\end{split}
\end{equation} for $t\in[t_{j},t_{j+1}),j\in\mathbb{N}$. It can be shown that the backstepping transformation \eqref{ctbt}, applied on the system \eqref{ctoe1},\eqref{ctoe2},\eqref{mctoe3} between $t_{j}$ and $t_{j+1}$, yields the following target system, valid for $t\in[t_{j},t_{j+1}),j\in\mathbb{N}$:
\begin{subequations}\label{ettsm}
\begin{align}
\hat{w}_{t}(x,t)&=\varepsilon \hat{w}_{xx}(x,t)+g(x)\tilde{w}(0,t),
\\
\hat{w}_{x}(0,t)&=-\frac{\lambda}{2\varepsilon}\tilde{w}(0,t),
\\
\hat{w}_{x}(1,t)&=-r\hat{w}(1,t)+d(t),
\end{align}
\end{subequations}
where $g(x)$ and $r$ are given by \eqref{gt} and \eqref{rt}, respectively.

It is straightforward to see that the observer error system $\tilde{u}$ for $t\in[t_{j},t_{j+1}),j\in\mathbb{N}$ under the modified boundary conditions \eqref{mctpe3} and \eqref{mctoe3} will still be the same as \eqref{ctoe}. Therefore, the application of  the backstepping transformation \eqref{ctobt} on $\tilde{u}$ between $t_{j}$ and $t_{j+1}$ yields the following observer error target system
\begin{subequations}\label{etoet}
\begin{align}\label{etoet1}
\tilde{w}_{t}(x,t)&=\varepsilon \tilde{w}_{xx}(x,t),
\\\label{etoet2}
\tilde{w}_{x}(0,t)&=0,
\\\label{etoet3}
\tilde{w}_{x}(1,t)&=-q\tilde{w}(1,t),
\end{align}
\end{subequations}
valid for $t\in[t_{j},t_{j+1}),j\in\mathbb{N}$.

\subsection{Well-posedness Issues}

\begin{prop}\label{cor1}
For every given initial data $u[t_{j}],\hat{u}[t_{j}]\in L^{2}(0,1)$, there exist unique mappings $u,\hat{u}\in C^{0}([t_{j},t_{j+1}];L^{2}(0,1))\cap C^{1}((t_{j},t_{j+1})\times [0,1])$ with $u[t],\hat{u}[t]\in C^{2}([0,1])$ which satisfy \eqref{ctpe2},\eqref{ctoe2},\eqref{etcl}-\eqref{mctoe3} for $t\in (t_{j},t_{j+1}]$ and \eqref{ctpe1}, \eqref{ctoe1} for $t\in (t_{j},t_{j+1}]$, $x\in(0,1)$.    
\end{prop}

\textit{Proof:} The initial condition $\tilde{w}[t_{j}]$ for the system \eqref{etoet} can be uniquely determined by using \eqref{diff} and \eqref{ibtoe} once $u[t_{j}]$ and $\hat{u}[t_{j}]$ are given. Therefore, from the straightforward application of Theorem 4.11 in \cite{karafyllis2019input}, we can show that there exist unique mappings $u,\tilde{w}\in C^{0}([t_{j},t_{j+1}];L^{2}(0,1))\cap C^{1}((t_{j},t_{j+1})\times [0,1])$ with $u[t],\tilde{w}[t]\in C^{2}([0,1])$ which satisfy \eqref{ctpe2},\eqref{etcl},\eqref{mctpe3},\eqref{etoet2},\eqref{etoet3} for $t\in (t_{j},t_{j+1}]$ and \eqref{ctpe1},\eqref{etoet1} for $t\in (t_{j},t_{j+1}]$, $x\in(0,1)$.  
Further, due to \eqref{diff} and the transformation \eqref{ctobt}, there also exists a unique mapping $\hat{u}\in C^{0}([t_{j},t_{j+1}];L^{2}(0,1))\cap C^{1}((t_{j},t_{j+1})\times [0,1])$ with $\hat{u}[t]\in C^{2}([0,1])$ which satisfy \eqref{ctoe2},\eqref{etcl},\eqref{mctoe3} for $t\in (t_{j},t_{j+1}]$and \eqref{ctoe1} for $t\in (t_{j},t_{j+1}]$, $x\in(0,1)$.$\qed$

\section{Observer-based Event-triggered Boundary Control}
\begin{figure}
\centering
\includegraphics[scale=1]{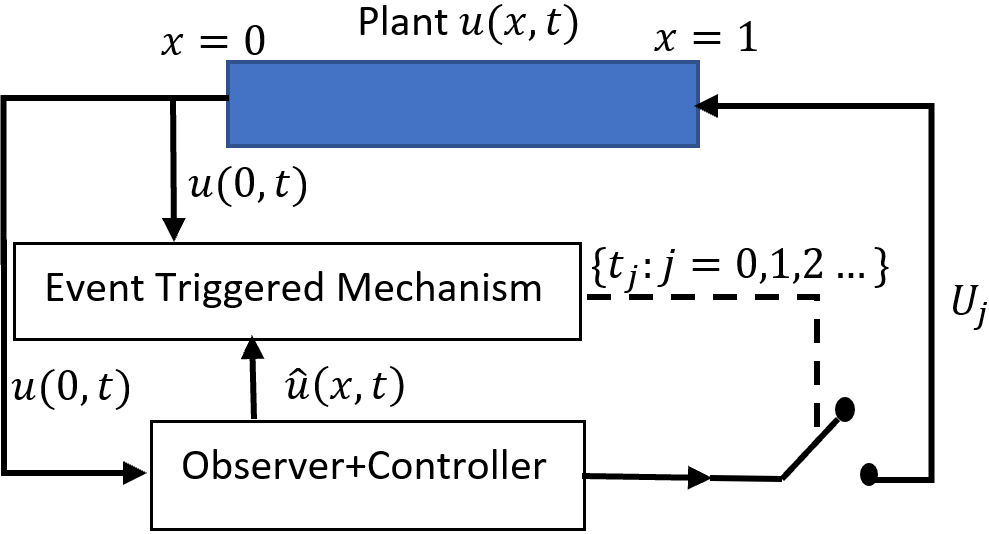}
\caption{Event-triggered observer-based closed-loop system.}
\end{figure}
Let us now present the observer-based event-triggered boundary control approach considered in this work. It consists of two components: 1) an event-triggered mechanism which decides the time instants at which the control value needs to be sampled/updated and 2) the observer-based backstepping output feedback controller. The structure of the closed-loop
system consisting of the plant, the observer-based controller, and the event trigger is illustrated in Fig. 1. The event-triggering condition is based on the square of the input holding error $d(t)$ and a dynamic variable $m(t)$ which depends on the information of the systems \eqref{ettsm} and \eqref{etoet}.\smallskip

\begin{dfn}\label{def1}
Let $\eta,\rho,\beta_{1},\beta_{2},\beta_{3}>0$. The observer-based event-triggered boundary control strategy consists of two components.
\begin{enumerate}
\item (The event-trigger) For some $j\in\mathbb{N}$, the event times $t_{j}\geq 0$ with $t_{0}=0$ form a finite increasing sequence via the following rules: \begin{itemize}
\item if $\{t\in\mathbb{R}_{+}|t>t_{j}\wedge  d^{2}(t)> -m(t)\}=\emptyset$ then the set of the times of the events is $\{t_{0},\ldots,t_{j}\}.$
\item if $\{t\in\mathbb{R}_{+}|t>t_{j}\wedge d^{2}(t)> -m(t)\}\neq\emptyset$ then the next event time is given by:\begin{equation}\label{obetbc1}
t_{j+1}=\inf\{t\in\mathbb{R}_{+}|t>t_{j}\wedge d^{2}(t)>-m(t)\},
\end{equation} where $d(t)$ is given by\begin{equation}\label{obetbc2}
\begin{split}
d(t)=\int_{0}^{1}\Big(&rK(1,y)+K_{x}(1,y)\Big)\\&\times\big(\hat{u}(y,t_{j})-\hat{u}(y,t)\big)dy,
\end{split}
\end{equation} for all $t\in[t_{j},t_{j+1})$ and $m(t)$ satisfies the ODE 
\begin{equation}\label{obetbc3}\begin{split}
\dot{m}(t)=&-\eta m(t)+\rho d^{2}(t)-\beta_{1}\Vert\hat{w}[t]\Vert^{2}\\&-\beta_{2}\vert\hat{w}(1,t)\vert^{2}-\beta_{3}\vert\tilde{w}(0,t)\vert^{2},\end{split}
\end{equation}  for all $t\in(t_{j},t_{j+1})$ with $m(t_{0})=m(0)<0$ and $m(t_{j}^{-})=m(t_{j})=m(t_{j}^{+})$.  
\end{itemize}
\item (The control action) The output boundary feedback control law\begin{equation}\label{obetbc4}
U_{j}=\int_{0}^{1}\Big(rK(1,y)+K_{x}(1,y)\Big)\hat{u}(y,t_{j})dy,
\end{equation}for all $t\in[t_{j},t_{j+1}),j\in\mathbb{N}$.
\end{enumerate}
\end{dfn}\smallskip

In Definition \ref{def1}, it is worth noting that the initial condition for  $m(t)$ in each time interval has been chosen such that $m(t)$ is time-continuous.

Proposition \eqref{cor1} allows us to define the solution of the closed-loop system under the observer-based event-triggered boundary control \eqref{obetbc1}-\eqref{obetbc4} in the interval $[0,F)$, where $F=\lim_{j\rightarrow\infty} (t_j)$.\smallskip 

\begin{lem}\label{lem1}Under the definition of the observer-based event-triggered boundary control \eqref{obetbc1}-\eqref{obetbc4}, it holds that $d^{2}(t)\leq-m(t)$ and $m(t)< 0,$ for $t\in [0,F)$ where $F=\lim_{j\rightarrow\infty} (t_j)$.
\end{lem}
\textit{Proof:} From Definition \ref{def1}, the events are triggered to guarantee $d^{2}(t)\leq-m(t),t\in [0,F)$. This inequality in combination with \eqref{obetbc3} yields:  
\begin{equation}\label{lem1e1}
\begin{split}
\dot{m}(t)\leq-(\eta+\rho) m(t)&-\beta_{1}\Vert\hat{w}[t]\Vert^{2}\\&-\beta_{2}\vert\hat{w}(1,t)\vert^{2}-\beta_{3}\vert\tilde{w}(0,t)\vert^{2},
\end{split}
\end{equation}for $ t\in(t_{j},t_{j+1}),j\in\mathbb{N}.$ Thus, considering the time-continuity of $m(t)$, we can obtain the following estimate:
\begin{equation}\label{fa}
\begin{split}
&m(t)\leq m(t_{j})e^{-(\eta+\rho)(t-t_{j})}\\&-\int_{t_{j}}^{t}e^{-(\eta+\rho)(t-\tau)}\big(\beta_{1}\Vert\hat{w}[\tau]\Vert^{2}\\&\hspace{30pt}+\beta_{2}\vert\hat{w}(1,\tau)\vert^{2}+\beta_{3}\vert\tilde{w}(0,\tau)\vert^{2}\big)d\tau,
\end{split}
\end{equation}
for $t\in[t_{j},t_{j+1}],j\in\mathbb{N}$. From Definition \ref{def1}, we have that \mbox{$m(t_{0})=m(0)<0$}. Therefore, it follows from \eqref{fa} that $m(t)<0$  for all $t\in[0,t_{1}]$. Again using \eqref{fa} on $[t_{1},t_{2}]$, we can show that $m(t)< 0$  for all $t\in[t_{1},t_{2}]$. Applying the same reasoning successively to the future intervals, it can be shown that $m(t)< 0$ for $t\in [0,F)$.$\qed$\smallskip

\begin{lem}\label{lem2} For $d(t)$ given by \eqref{obetbc2}, it holds that\begin{equation}\label{ghm}
\dot{d}^{2}(t)\leq \rho_{1} d^{2}(t)+\alpha_{1}\Vert\hat{w}[t]\Vert^{2}+\alpha_{2}\vert\hat{w}(1,t)\vert^{2}+\alpha_{3}\vert\tilde{w}(0,t)\vert^{2},
\end{equation}
for some $\rho_{1},\alpha_{1},\alpha_{2},\alpha_{3}>0,$ for all $t\in(t_{j},t_{j+1}),j\in\mathbb{N}.$\end{lem}

\textit{Proof:} From \eqref{obetbc2}, we can show for $t\in(t_{j},t_{j+1}),j\in\mathbb{N}$\begin{equation}\label{dad}
\dot{d}(t)=-\int_{0}^{1}k(y)\hat{u}_{t}(y,t)dy,
\end{equation}
where
\begin{equation}\label{frk}
k(y)=rK(1,y)+K_{x}(1,y).
\end{equation}Using \eqref{ctoe1} on \eqref{dad} and integrating by parts twice in the interval $(t_{j},t_{j+1}),j\in\mathbb{N}$, we can show that 
\begin{equation}
\begin{split}
&\dot{d}(t)=-\varepsilon\int_{0}^{1}k(y)\hat{u}_{yy}(y,t)dy-\lambda\int_{0}^{1}k(y)\hat{u}(y,t)dy\\&\hspace{28pt}-\int_{0}^{1}k(y)p_{1}(y)dy\tilde{u}(0,t)
\\&=-\varepsilon k(1)\hat{u}_{x}(1,t)+\varepsilon k(0)\hat{u}_{x}(0,t)+\varepsilon\frac{dk(x)}{dx}\Big |_{x=1}\hat{u}(1,t)\\&\hspace{12pt}-\varepsilon\frac{dk(x)}{dx}\Big|_{x=0}\hat{u}(0,t)
-\varepsilon\int_{0}^{1}\frac{d^{2}k(y)}{dy^{2}}\hat{u}(y,t)dy\\&\hspace{12pt}-\lambda\int_{0}^{1}k(y)\hat{u}(y,t)dy-\int_{0}^{1}k(y)p_{1}(y)dy\tilde{u}(0,t).
\end{split}
\end{equation}Furthermore, using \eqref{ctoe2},\eqref{mctoe3}, \eqref{obetbc2}, and \eqref{obetbc4}, we can show that 
\begin{equation}\label{ddt}
\begin{split}
\dot{d}(t)=&-\varepsilon k(1)d(t)+\Big(\varepsilon qk(1)+\varepsilon\frac{dk(x)}{dx}\Big\vert_{x=1}\Big)\hat{u}(1,t)\\&-\int_{0}^{1}\Big(\varepsilon\frac{d^{2}k(y)}{dy^{2}}+\varepsilon k(1)k(y)+\lambda k(y)\Big)\hat{u}(y,t)dy\\&-\Big(\frac{\lambda k(0)}{2}+\int_{0}^{1}k(y)p_{1}(y)dy\Big)\tilde{u}(0,t).
\end{split}
\end{equation}
It is worth mentioning that above we have used the fact that $dk(x)/dx\Big\vert_{x=0}=0,$ which can be shown using \eqref{frk} and \eqref{ctcks}. Using Young's and Cauchy-Schwarz inequalities on \eqref{ink},  we also can show that\begin{equation}\label{yc1}
\Vert\hat{u}[t]\Vert^{2}\leq \bigg(1+\Big(\int_{0}^{1}\int_{0}^{x}L^{2}(x,y)dydx\Big)^{1/2}\bigg)^{2}\Vert\hat{w}[t]\Vert^{2},
\end{equation}\begin{equation}\label{yc2}
\hat{u}^{2}(1,t)\leq 2\hat{w}^{2}(1,t)+2\int_{0}^{1}L^{2}(1,y)dy\Vert\hat{w}[t]\Vert^{2}.
\end{equation}
Using Young's and Cauchy-Schwarz  inequalities repeatedly on \eqref{ddt} along with \eqref{yc1} and \eqref{yc2}, we can show that
\begin{equation}
\begin{split}
\dot{d}^{2}(t)\leq& \rho_{1} d^{2}(t)+\alpha_{1}\Vert\hat{w}[t]\Vert^{2}+\alpha_{2}\vert\hat{w}(1,t)\vert^{2}\\&+\alpha_{3}\vert\tilde{w}(0,t)\vert^{2},
\end{split}
\end{equation}
where\begin{align}
\label{neqro}
\rho_{1}&=6\varepsilon^{2}k^{2}(1),\\
\begin{split}
\label{al1}\alpha_{1}&=3\bigg(1+\Big(\int_{0}^{1}\int_{0}^{x}L^{2}(x,y)dydx\Big)^{1/2}\bigg)^{2}\\&\quad\times\int_{0}^{1}\Big(\varepsilon\frac{d^{2}k(y)}{dy^{2}}+\varepsilon k(1)k(y)+\lambda k(y)\Big)^{2}dy\\&\quad+6\Big(\varepsilon qk(1)+\varepsilon\frac{dk(x)}{dx}\Big\vert_{x=1}\Big)^{2}\int_{0}^{1}L^{2}(1,y)dy,
\end{split}\\
\label{al2}
\alpha_{2}&=6\Big(\varepsilon qk(1)+\varepsilon\frac{dk(x)}{dx}\Big\vert_{x=1}\Big)^{2},\\
\label{al3}
\alpha_{3}&=6\Big(\frac{\lambda k(0)}{2}+\int_{0}^{1}k(y)p_{1}(y)dy\Big)^{2}.
\end{align}
$\qed$

\section{Main Results}

\begin{thme}Under the observer-based event-triggered boundary control in Definition \ref{def1}, with $\beta_{1},\beta_{2},\beta_{3}$ chosen as\begin{equation}\label{betas}
\beta_{1}=\alpha_{1}/(1-\sigma),\hspace{5pt}\beta_{2}=\alpha_{2}/(1-\sigma),\hspace{5pt}\beta_{3}=\alpha_{3}/(1-\sigma),
\end{equation}where $\alpha_{1},\alpha_{2},\alpha_{3}$ given by \eqref{al1}-\eqref{al3} and $\sigma\in(0,1)$,  there exists a minimal dwell-time $\tau>0$ between two triggering times, \textit{i.e.,} there exists a constant $\tau>0$ such that $t_{j+1}-t_{j}\geq\tau,$ for all $j\in\mathbb{N}$, which is independent of the initial conditions and only depends on the system and control parameters.\end{thme}

\textit{Proof:} From Lemma \ref{lem1}, we have that  $d^{2}(t)\leq -(1-\sigma)m(t)-\sigma m(t),$ where $\sigma\in(0,1)$ and $m(t)<0$ for $t\in [0,F)$, where $F=\lim_{j\rightarrow\infty} (t_j)$.  Let us define the function\begin{equation}\label{fmdt}
\psi(t):=\frac{ d^{2}(t)+(1-\sigma)m(t)}{-\sigma m(t)}.
\end{equation}
Note that $\psi(t)$ is continuous in $[t_{j},t_{j+1})$. A lower bound for the inter-execution times is given by the time it takes for the function $\psi$ to go from $\psi(t_{j})$ to $\psi(t_{j+1}^{-})=1,$ where $\psi(t_{j})<0$, which holds since $d(t_{j})=0$. Here $t_{j+1}^{-}$ is the left limit at $t=t_{j+1}$.  So, by the intermediate value theorem, there exists a $t_{j}^{'}>t_{j}$ such that $\psi(t'_{j})=0$ and $\psi(t)\in[0,1]$ for $t\in[t_{j}^{'},t_{j+1}^{-}]$. The time derivative of $\psi$ on $[t_{j}^{'},t_{j+1})$ is given by
\begin{equation}
\dot{\psi}(t)=\frac{2d(t)\dot{d}(t)+(1-\sigma)\dot{m}(t)}{-\sigma m(t)}-\frac{\dot{m}(t)}{m(t)}\psi(t).
\end{equation}
From Young's inequality, we have that
\begin{equation}
\dot{\psi}(t)\leq\frac{d^{2}(t)+\dot{d}^{2}(t)+(1-\sigma)\dot{m}(t)}{-\sigma m(t)}-\frac{\dot{m}(t)}{m(t)}\psi(t).
\end{equation}
Using Lemma \ref{lem2} and \eqref{obetbc3}, we can show that
\begin{equation}\label{met1}
\begin{split}
&\dot{\psi}(t)\leq\frac{\Big(1+\rho_{1}+(1-\sigma)\rho\Big)d^{2}(t)}{-\sigma m(t)}+\frac{(1-\sigma)\eta}{\sigma}\\&+\eta\psi(t)+\frac{\sigma \rho d^{2}(t)}{-\sigma m(t)}\psi(t)
+\frac{\big(\alpha_{1}-(1-\sigma)\beta_{1}\big)\Vert\hat{w}[t]\Vert^{2}}{-\sigma m(t)}\\&+\frac{\big(\alpha_{2}-(1-\sigma)\beta_{2}\big)\vert\hat{w}(1,t)\vert^{2}}{-\sigma m(t)}\\&+\frac{\big(\alpha_{3}-(1-\sigma)\beta_{3}\big)\vert\tilde{w}(0,t)\vert^{2}}{-\sigma m(t)}\\&
+\frac{\beta_{1}\Vert\hat{w}[t]\Vert^{2}+\beta_{2}\vert\hat{w}(1,t)\vert^{2}+\beta_{3}\vert \tilde{w}(0,t)\vert^{2}}{m(t)}\psi(t).
\end{split}
\end{equation}Let us choose $\beta_{1},\beta_{2},\beta_{3}$ as in \eqref{betas}, where $\alpha_{1},\alpha_{2},\alpha_{3}$ are given by \eqref{al1}-\eqref{al3}, respectively.  Also, note that the last term in the right hand side of \eqref{met1} is negative. Therefore, we have
\begin{equation}
\begin{split}
\dot{\psi}(t)\leq& \frac{\Big(1+\rho_{1}+(1-\sigma)\rho\Big)d^{2}(t)}{-\sigma m(t)}+\frac{(1-\sigma)\eta}{\sigma}\\&+\eta\psi(t)+\frac{\sigma \rho d^{2}(t)}{-\sigma m(t)}\psi(t).
\end{split}
\end{equation}
We also can write that
\begin{equation}\label{zbe}
\begin{split}
&\dot{\psi}(t)\leq\Big(1+\rho_{1}+(1-\sigma)\rho\Big)\frac{\Big( d^{2}(t)+(1-\sigma)m(t)\Big)}{-\sigma m(t)}\\&+\frac{(1-\sigma)\eta}{\sigma}+\eta \psi(t)+\sigma\rho\frac{d^{2}(t)+(1-\sigma) m(t)}{-\sigma m(t)}\psi(t)\\&+\Big(1+\rho_{1}+(1-\sigma)\rho\Big)\frac{(1-\sigma)}{\sigma}+\rho(1-\sigma)\psi(t).
\end{split}
\end{equation}
We can rewrite \eqref{zbe} as
\begin{equation}
\dot{\psi}(t)\leq a_{1}\psi^{2}(t)+a_{2}\psi(t)+a_{3},
\end{equation}
where \begin{align}
\label{a1}a_{1}&=\sigma\rho>0,\\
\label{a2}a_{2}&=1+\rho_{1}+2(1-\sigma)\rho+\eta>0,\\
\label{a3}a_{3}&=\big(1+\rho_{1}+(1-\sigma)\rho+\eta\big)\frac{1-\sigma}{\sigma}>0.
\end{align}By the Comparison principle, it follows that the time needed for $\psi$ to go from $\psi(t_{j}^{'})=0$ to $\psi(t_{j+1}^{-})=1$ is at least 
\begin{equation}\label{mdt}
\tau=\int_{0}^{1}\frac{1}{a_{1}s^{2}+a_{2}s+a_{3}}ds.
\end{equation}Therefore, $t_{j+1}-t_{j}^{'}\geq\tau$. As $t_{j+1}-t_{j}\geq t_{j+1}-t_{j}^{'}$, we can conclude that $t_{j+1}-t_{j}\geq\tau$. Thus, $\tau$ can be considered a lower bound for the minimal dwell-time. Note that $\tau$ is independent of initial conditions and only depends on system and control parameters. $\qed$\smallskip

\begin{cllry}\label{corf}
For every given initial data $u[0],\hat{u}[0]\in L^{2}(0,1)$, there exist unique mappings \mbox{$u,\hat{u}\in C^{0}(\mathbb{R_{+}};L^{2}(0,1))\cap C^{1}(I\times [0,1])$} with $u[t],\hat{u}[t]\in C^{2}([0,1])$ which satisfy \eqref{ctpe2},\eqref{ctoe2},\eqref{etcl}-\eqref{mctoe3} for all $t>0$ and \eqref{ctpe1}, \eqref{ctoe1} for all $t>0$, $x\in(0,1)$, where $I=\mathbb{R_{+}}\text{\textbackslash}\{t_{j}\geq 0,j\in\mathbb{N}\}$. The increasing sequence $\{t_{j}\geq 0,j\in\mathbb{N}\}$ is determined by the set of rules  given in Definition \ref{def1}.
\end{cllry}

\textit{Proof:} This is a straightforward consequence of Proposition \ref{cor1} and Theorem 4.10 in \cite{karafyllis2019input}. The solutions are constructed iteratively between consecutive triggering times.$\qed$\smallskip

\begin{thme}\label{thm2}Let $\eta>0$  be a design parameter, $\sigma\in(0,1)$, and $g(x)$ and $r$ be given by \eqref{gt} and \eqref{rt}, respectively, while $\beta_{1},\beta_{2},\beta_3$ are chosen according to \eqref{betas}. Further, subject to Assumption \ref{ass1}, let us choose parameters  $B,\kappa_{1},\kappa_{2},\kappa_{3}>0$ such that
\begin{equation}\label{sapie1}
\begin{split}
B\bigg(\varepsilon\min\Big\{r-\frac{1}{2},\frac{1}{2}\Big\}-\frac{\varepsilon}{2\kappa_{1}}&-\frac{5\lambda}{8\kappa_{2}}-\frac{\Vert g\Vert^{2}}{\kappa_{3}}\bigg)\\&-2\beta_{1}-\beta_{2}>0,
\end{split}
\end{equation}$A$ such that
\begin{equation}\label{sapie2}
A\varepsilon\min\Big\{q-\frac{1}{2},\frac{1}{2}\Big\}-\frac{5\lambda \kappa_{2}B}{8}-\frac{5\kappa_{3}B}{4}-\frac{5\beta_{3}}{2}>0,
\end{equation}and $\rho$ as
\begin{equation}\label{thmrho}
\rho=\frac{\varepsilon\kappa_{1} B}{2}.
\end{equation}
Then, the closed-loop system which consists of the plant  \eqref{ctpe1},\eqref{ctpe2},\eqref{mctpe3} and  the observer \eqref{ctoe1},\eqref{ctoe2},\eqref{mctoe3} with the  event-triggered boundary controller \eqref{obetbc1}-\eqref{obetbc4} has a unique solution and globally exponentially converges to zero, \text{i.e.,} $\Vert u[t]\Vert+\Vert\hat{u}[t]\Vert\rightarrow 0$ as $t\rightarrow \infty.$\end{thme}

\textit{Proof:} From Corollary \ref{corf}, the existence and the uniqueness of solutions
to the plant \eqref{ctpe1},\eqref{ctpe2},\eqref{mctpe3} and the observer \eqref{ctoe1},\eqref{ctoe2},\eqref{mctoe3} are guaranteed.  Now let us show that the closed-loop system is globally $L^{2}$-exponentially convergent to zero.

Let us choose the following candidate Lyapunov function noting  that $m(t)<0$ for $t\geq 0$:\begin{equation}\label{lpnv}
V=\frac{A}{2}\int_{0}^{1}\tilde{w}^{2}(x,t)dx+\frac{B}{2}\int_{0}^{1}\hat{w}^{2}(x,t)dx-m(t).
\end{equation}
Here $\hat{w}$ and $\tilde{w}$ are the systems described by \eqref{ettsm} and \eqref{etoet}, respectively.  We can show that for $t\in(t_{j},t_{j+1}),j\in\mathbb{N}$ 
\begin{equation}\label{dlf}
\begin{split}
\dot{V}=&-A\varepsilon q\tilde{w}^{2}(1,t)-A\varepsilon\int_{0}^{1}\tilde{w}^{2}_{x}(x,t)dx\\&-r\varepsilon B\hat{w}^{2}(1,t)+\varepsilon B d(t)\hat{w}(1,t)+\frac{\lambda B}{2}\hat{w}(0,t)\tilde{w}(0,t)\\&-\varepsilon B\int_{0}^{1}\hat{w}_{x}^{2}(x,t)dx+B\int_{0}^{1}g(x)\hat{w}(x,t)dx\tilde{w}(0,t)\\&-\dot{m}(t).
\end{split}
\end{equation}
From Young's and Cauchy-Schwarz inequalities, we can write that
\begin{equation}\label{ie1}
\varepsilon B d(t)\hat{w}(1,t)\leq \frac{\varepsilon B}{2\kappa_{1}}\hat{w}^{2}(1,t)+\frac{\varepsilon \kappa_{1}B}{2}d^{2}(t),
\end{equation}\begin{equation}\label{ie2}
\frac{\lambda B}{2}\hat{w}(0,t)\tilde{w}(0,t)\leq\frac{\lambda B}{4\kappa_{2}}\hat{w}^{2}(0,t)+\frac{\lambda\kappa_{2}B}{4}\tilde{w}^{2}(0,t),
\end{equation}\begin{equation}\label{ie3}
\begin{split}
B\int_{0}^{1}g(x)\hat{w}(x,t)dx\tilde{w}(0,t)\leq&\frac{\Vert g\Vert^{2}B}{2\kappa_{3}}\Vert \hat{w}[t]\Vert^{2}+\frac{\kappa_{3}B}{2}\tilde{w}^{2}(0,t).
\end{split}
\end{equation}Therefore, using \eqref{ie1}-\eqref{ie3},\eqref{obetbc3},\eqref{thmrho}, we can write \eqref{dlf} as
\begin{equation}\label{lfe1}
\begin{split}
\dot{V}\leq&-A\varepsilon q\tilde{w}^{2}(1,t)-A\varepsilon\Vert\tilde{w}_{x}[t]\Vert^{2}-r\varepsilon B\hat{w}^{2}(1,t)\\&+\frac{\varepsilon B}{2\kappa_{1}}\hat{w}^{2}(1,t)+\frac{\lambda B}{4\kappa_{2}}\hat{w}^{2}(0,t)+\frac{\lambda\kappa_{2} B}{4}\tilde{w}^{2}(0,t)\\&
-\varepsilon B\Vert\hat{w}_{x}[t]\Vert^{2}+\frac{\Vert g\Vert^{2}B}{2\kappa_{3}}\Vert\hat{w}[t]\Vert^{2}+\frac{\kappa_{3} B}{2}\tilde{w}^{2}(0,t)\\&+\beta_{1}\Vert\hat{w}[t]\Vert^{2}+\beta_{2}\hat{w}^{2}(1,t)\\&+\beta_{3}\tilde{w}^{2}(0,t)+\eta m(t).
\end{split}
\end{equation}From Agmon's and Young's inequalities, we have that\begin{align}
\tilde{w}^{2}(0,t)&\leq \tilde{w}^{2}(1,t)+\Vert \tilde{w}[t]\Vert^{2}+\Vert\tilde{w}_{x}[t]\Vert^{2},\\
\hat{w}^{2}(0,t)&\leq \hat{w}^{2}(1,t)+\Vert \hat{w}[t]\Vert^{2}+\Vert\hat{w}_{x}[t]\Vert^{2}.
\end{align}Therefore, we can show from \eqref{lfe1} that
\begin{equation}\label{aa}
\begin{split}
\dot{V}\leq& -\Big(A\varepsilon q-\frac{\lambda\kappa_{2}B}{4}-\frac{\kappa_{3}B}{2}-\beta_{3}\Big)\tilde{w}^{2}(1,t)\\&-\Big(A\varepsilon-\frac{\lambda\kappa_{2}B}{4}-\frac{\kappa_{3}B}{2}-\beta_{3}\Big)\Vert\tilde{w}_{x}[t]\Vert^{2}\\&+\Big(\frac{\lambda\kappa_{2}B}{4}+\frac{\kappa_{3}B}{2}+\beta_{3}\Big)\Vert\tilde{w}[t]\Vert^{2}\\&
-\Big(r\varepsilon B-\frac{\varepsilon B}{2\kappa_{1}}-\frac{\lambda B}{4\kappa_{2}}-\beta_{2}\Big)\hat{w}^{2}(1,t)\\&-\Big(\varepsilon B-\frac{\lambda B}{4\kappa_{2}}\Big)\Vert\hat{w}_{x}[t]\Vert^{2}\\&+\Big(\frac{\lambda B}{4\kappa_{2}}+\frac{\Vert g\Vert^{2}B}{2\kappa_{3}}+\beta_{1}\Big)\Vert\hat{w}[t]\Vert^{2}+\eta m(t).
\end{split}
\end{equation}
From Poincar\'e Inequality, we have that\begin{align}
\label{aa1}-\Vert \tilde{w}_{x}[t]\Vert^{2}&\leq \frac{1}{2}\tilde{w}^{2}(1,t)-\frac{1}{4}\Vert \tilde{w}[t]\Vert^{2},\\
\label{aa2}-\Vert \hat{w}_{x}[t]\Vert^{2}&\leq \frac{1}{2}\hat{w}^{2}(1,t)-\frac{1}{4}\Vert \hat{w}[t]\Vert^{2}.
\end{align}Furthermore, we have from \eqref{sapie1}  and \eqref{sapie2} that\begin{equation}\label{aa3}
A\varepsilon-\frac{\lambda\kappa_{2} B}{4}-\frac{\kappa_{3}B}{2}-\beta_{3}>0,\text{ and } \varepsilon B-\frac{\lambda B}{4\kappa_{2}}>0.
\end{equation}Therefore, using \eqref{aa}-\eqref{aa3}, we can show that
\begin{equation}\label{bb}
\begin{split}
\dot{V}\leq &-\Big(A\varepsilon(q-\frac{1}{2})-\frac{\lambda\kappa_{2} B}{8}-\frac{\kappa_{3} B}{4}-\frac{\beta_{3}}{2}\Big)\tilde{w}^{2}(1,t)\\&-\Big(\frac{A\varepsilon}{4}-\frac{5\lambda\kappa_{2}B}{16}-\frac{5\kappa_{3}B}{8}-\frac{5\beta_{3}}{4}\Big)\Vert\tilde{w}[t]\Vert^{2}\\&
-\Big(\varepsilon B(r-\frac{1}{2})-\frac{\varepsilon B}{2\kappa_{1}}-\frac{\lambda B}{8\kappa_{2}}-\beta_{2}\Big)\hat{w}^{2}(1,t)\\&-\Big(\frac{\varepsilon B}{4}-\frac{5\lambda B}{16\kappa_{2}}-\frac{\Vert g\Vert^{2}B}{2\kappa_{3}}-\beta_{1}\Big)\Vert\hat{w}[t]\Vert^{2}+\eta m(t).
\end{split}
\end{equation}
From \eqref{sapie1}  and \eqref{sapie2}, we can show that
\begin{align}
\label{bb1}A\varepsilon(q-\frac{1}{2})-\frac{\lambda\kappa_{2}B}{8}-\frac{\kappa_{3}B}{4}-\frac{\beta_{3}}{2}&>0,\\
\label{bb2}\varepsilon B(r-\frac{1}{2})-\frac{\varepsilon B}{2\kappa_{1}}-\frac{\lambda B}{8\kappa_{2}}-\beta_{2}&>0.
\end{align}Thus, using \eqref{bb}-\eqref{bb2}, we can obtain that

\begin{equation}
\begin{split}
&\dot{V}\leq-\Big(\frac{A\varepsilon}{4}-\frac{5\lambda\kappa_{2}B}{16}-\frac{5\kappa_{3}B}{8}-\frac{5\beta_{3}}{4}\Big)\Vert\tilde{w}[t]\Vert^{2}\\&-\Big(\frac{\varepsilon B}{4}-\frac{5\lambda B}{16\kappa_{2}}-\frac{\Vert g\Vert^{2}B}{2\kappa_{3}}-\beta_{1}\Big)\Vert\hat{w}[t]\Vert^{2}+\eta m(t).
\end{split}
\end{equation}Again, we have from \eqref{sapie1}  and \eqref{sapie2} that

\begin{align}
b_{1}&=\frac{A\varepsilon}{4}-\frac{5\lambda\kappa_{2}B}{16}-\frac{5\kappa_{3}B}{8}-\frac{5\beta_{3}}{4}>0,\\
b_{2}&=\frac{\varepsilon B}{4}-\frac{5\lambda B}{16\kappa_{2}}-\frac{\Vert g\Vert^{2}B}{2\kappa_{3}}-\beta_{1}>0.
\end{align}Therefore, we have for $t\in(t_{j},t_{j+1}),j\in\mathbb{N}$ that 

\begin{equation}\label{lst}
\dot{V}\leq-\varrho V,
\end{equation}
where
\begin{equation}
 \varrho=\frac{\min\big\{b_{1},b_{2},\eta\big\}}{\max\big\{A/2,B/2,1\big\}}.
\end{equation}
Concentrating on this time interval, we can show that $V(t^{-}_{j+1})\leq e^{-\varrho(t_{j+1}^{-}-t_{j}^{+})}V(t^{+}_{j})$.  Here $t_{j}^{+}$ and $t_{j}^{-}$ are the right and left limits of $t=t_{j}$. Since $V(t)$ is continuous (as $m(t),\Vert\hat{w}[t]\Vert$,$\Vert\tilde{w}[t]\Vert$ are continuous), we have that $V(t^{-}_{j+1})=V(t_{j+1})$ and $V(t_{j}^{+})=V(t_{j}),$ and therefore, 
\begin{equation}
V(t_{j+1})\leq e^{-\varrho(t_{j+1}-t_{j})}V(t_{j}).
\end{equation}
Hence, for any $t\geq 0$ in $t\in[t_{j},t_{j+1}),j\in\mathbb{N},$ we can also obtain the following:
\begin{equation}\label{ffr}
\begin{split}
V(t)&\leq e^{-\varrho(t-t_{j})}V(t_{j})\\&
\leq e^{-\varrho(t-t_{j})}\times e^{-\varrho(t_{j}-t_{j-1})}V(t_{j-1})\\&\leq\cdots\leq\\&\leq e^{-\varrho(t-t_{j})}\times
\prod_{i=1}^{i=j}e^{-\varrho(t_{i}-t_{i-1})} V(0)\\&=
e^{-\varrho t}V(0).
\end{split}
\end{equation}
Thus, recalling that $m(0)<0$ from Definition \ref{def1}, we have that\begin{equation}
\begin{split}
\frac{A}{2}\Vert\tilde{w}[t]\Vert^{2}&+\frac{B}{2}\Vert\hat{w}[t]\Vert^{2}-m(t)\\&\leq e^{-\varrho t}\Big(\frac{A}{2}\Vert\tilde{w}[0]\Vert^{2}+\frac{B}{2}\Vert\hat{w}[0]\Vert^{2}-m(0)\Big).
\end{split}
\end{equation}As $m(t)< 0$, we also have that 
\begin{equation}\label{crse}
\begin{split}
\frac{A}{2}\Vert\tilde{w}[t]\Vert^{2}&+\frac{B}{2}\Vert\hat{w}[t]\Vert^{2}\\&\leq e^{-\varrho t}\Big(\frac{A}{2}\Vert\tilde{w}[0]\Vert^{2}+\frac{B}{2}\Vert\hat{w}[0]\Vert^{2}-m(0)\Big).
\end{split}
\end{equation}This implies that the systems $\hat{w}$ and $\tilde{w}$ given by \eqref{ettsm} and \eqref{etoet}, respectively, are globally exponentially convergent in $L^{2}$-sense to zero. Using \eqref{ctobt} and \eqref{ink}, we can show that $\Vert\tilde{u}[t]\Vert^{2}\leq\tilde{P}^2\Vert\tilde{w}[t]\Vert^{2}$ and $\Vert \hat{u}[t]\Vert^{2}\leq \tilde{L}^2\Vert\hat{w}[t]\Vert^{2}$, respectively. Here $\tilde{P}=\bigg(1+\Big(\int_{0}^{1}\int_{0}^{x}P^{2}(x,y)dydx\Big)^{1/2}\bigg)^{2}$ and $\tilde{L}=\bigg(1+\Big(\int_{0}^{1}\int_{0}^{x}L^{2}(x,y)dydx\Big)^{1/2}\bigg)^{2}$.  Further from \eqref{diff}, we can show that $\Vert u[t]\Vert^{2}\leq2\Vert \tilde{u}[t]\Vert^{2}+2\Vert\hat{u}[t]\Vert^{2}$. Therefore, we can obtain from \eqref{crse} that

\begin{equation}
\min\Big\{\frac{A}{\tilde{P}},\frac{B}{\tilde{L}}\Big\}\Vert u[t]\Vert^{2}\leq 4e^{-\varrho t}\Big(\frac{A}{2}\Vert\tilde{w}[0]\Vert^{2}+\frac{B}{2}\Vert\hat{w}[0]\Vert^{2}-m(0)\Big),
\end{equation}
\begin{equation}
\frac{B}{\tilde{L}}\Vert\hat{u}[t]\Vert^2\leq 2e^{-\varrho t}\Big(\frac{A}{2}\Vert\tilde{w}[0]\Vert^{2}+\frac{B}{2}\Vert\hat{w}[0]\Vert^{2}-m(0)\Big).
\end{equation}
Therefore, we  can derive the following estimate:

\begin{equation}
\begin{split}
\Vert u[t]\Vert &+\Vert\hat{u}[t]\Vert\\&\leq \sqrt{\frac{12\Big(\frac{A}{2}\Vert\tilde{w}[0]\Vert^{2}+\frac{B}{2}\Vert\hat{w}[0]\Vert^{2}-m(0)\Big)}{\min\Big\{\frac{A}{\tilde{P}},\frac{B}{\tilde{L}}\Big\}}}e^{-\frac{\varrho t}{2}},
\end{split}
\end{equation}which implies that $\Vert u[t]\Vert +\Vert\hat{u}[t]\Vert\rightarrow 0$ as $t\rightarrow \infty$.$\qed$\smallskip
\begin{rmk}
In Theorem 2, we have established the global exponential convergence of the closed-loop system to the equilibrium point. It follows from \eqref{crse} that we could have obtained global exponential stability if we chose $m(0)=0$. However, if $m(0)=0$, then $m(t)\leq 0$ (this can be shown by following the same arguments in the proof of Lemma \ref{lem1}). Then, the function $\psi(t)$ in \eqref{fmdt} is not defined when $m(t)=0$. Therefore, the existence of a minimal-dwell time cannot be proved easily by following the same arguments as in the proof of Theorem 1. Hence, $m(0)$ has to be chosen strictly negative.  \smallskip 
\end{rmk}
\begin{rmk}
The parameter $\eta>0$ characterizes the decay rate of $m(t)$ governed by \eqref{obetbc3}. Thus, $\eta$ may be used to adjust the sampling speed of the event-triggered mechanism. The larger $\eta$, the faster is the sampling speed. We consider $\sigma\in(0,1)$ as a free parameter that can be tuned appropriately such that the conditions for guaranteeing a minimal dwell-time are met.
\smallskip
\end{rmk}

\begin{rmk}
We remark that if a periodic sampling scheme where the control value is periodically updated in a sampled-and-hold manner is to be used to stabilize the plant \eqref{ctp} and the observer \eqref{cto}, one can choose a sampling period $T$ upper bounded by the minimal dwell-time $\tau$ \eqref{mdt}. It will ensure the closed-loop system's global exponential convergence because the relation \eqref{ffr} is guaranteed to hold for all $T\leq \tau $. However, one should expect $\tau$ to be very small and conservative as the coefficients $a_{1},a_{2},$ and $a_{3}$ \eqref{a1}-\eqref{a3} are usually large. This issue, on the other hand, reinforces the motivation for ETC, that is sample and update only when required. 
\end{rmk}

\section{Numerical Simulations}

We consider a reaction-diffusion system with $\varepsilon=0.1;\lambda=0.25;q=2.3$ and the initial conditions $u[0]=10x^2(x-1)^{2}$ and $\hat{u}[0]=5x^2(x-1)^2+5x^3(x-1)^3$. For numerical simulations, both  plant and observer states are discretized with uniform step size of $h=0.0062$ for the space variable. Time discretization was done using the implicit Euler scheme with a step size $\Delta t=h$. 

The parameters for the event-trigger mechanism are chosen as follows: $m(0)=-10^{-4},\eta=1\text{ or }100$ and  $\sigma=0.1$. It can be shown using \eqref{al1}-\eqref{al3} that $\alpha_{1}=4.14;\alpha_2=2.07;\alpha_3=3.3$. Therefore, from \eqref{betas}, we can obtain $\beta_{1}=4.6;\beta_{2}=2.3;\beta_{3}=3.7$. Finding that $\Vert g\Vert^{2}= 0.0297$, let us choose $\kappa_1=2.1;\kappa_2=312.5;\kappa_3=59.4$ and $B=460$ to satisfy \eqref{sapie1}. Then, from \eqref{thmrho}, we can obtain $\rho=48.3$.

Fig. 2 shows the zero-input response of the plant and it is clear that the system is unstable. Fig. 3(a) shows the response of the pant under ETC with $\eta=1$  and Fig. 3(b) shows the evolution of $\Vert u[t]\Vert,\Vert \hat{u}[t]\Vert$ and $\Vert\tilde{u}[t]\Vert$. The evolution of the control input when $\eta=1$ is presented in Fig. 4(a) along with the corresponding continuous-time control input. The behavior of the functions associated with the triggering condition \eqref{obetbc1}  for the case of $\eta=1$ is depicted in Fig. 4(b). Once the trajectory $d^{2}(t)$ reaches the trajectory $-m(t)$, an event is generated, the control input is updated and $d(t)$ is reset to zero. Fig. 5 compares the ETC control input for $\eta=1$ and $\eta=100$, and it can be seen that $\eta=100$ results in faster sampling than $\eta=1$. 

Finally, we conduct simulations for $100$ different initial conditions $u[0]=x^2(x-1)^2\sin(n\pi x), n=1,\ldots,100$ and $\hat{u}[0]=2u[t]$ on a time frame of $150\text{ }s$. Next, we compute the inter-execution times between two events and compare the cases for slow and fast sampling, \textit{i.e.,} $\eta=1$ and $\eta=100$, respectively. Fig. 6 shows the density of the inter-execution times plotted in logarithmic scale, and it can be stated that when $\eta$ is smaller, the inter-executions times are larger and the sampling is less often. For $\eta=1$ and $\eta=100$, the inter-execution times are typically in the range $0.1\text{ }s-10\text{ }s$. For the system under consideration, the minimal dwell time $\tau$ calculated using \eqref{mdt} when $\eta=1$ and $\eta=100$ are respectively $2.2\times 10^{-3}\text{ }s$  and $7.2\times 10^{-4} \text{ }s$. Therefore, the fact that an analogous  sampled-data controller guaranteeing exponential convergence has to be implemented using these small and conservative sampled periods manifests the importance and the need of ETC. 

\begin{figure}\label{fig1}
\centering
\subfloat[]{\includegraphics[scale=.11]{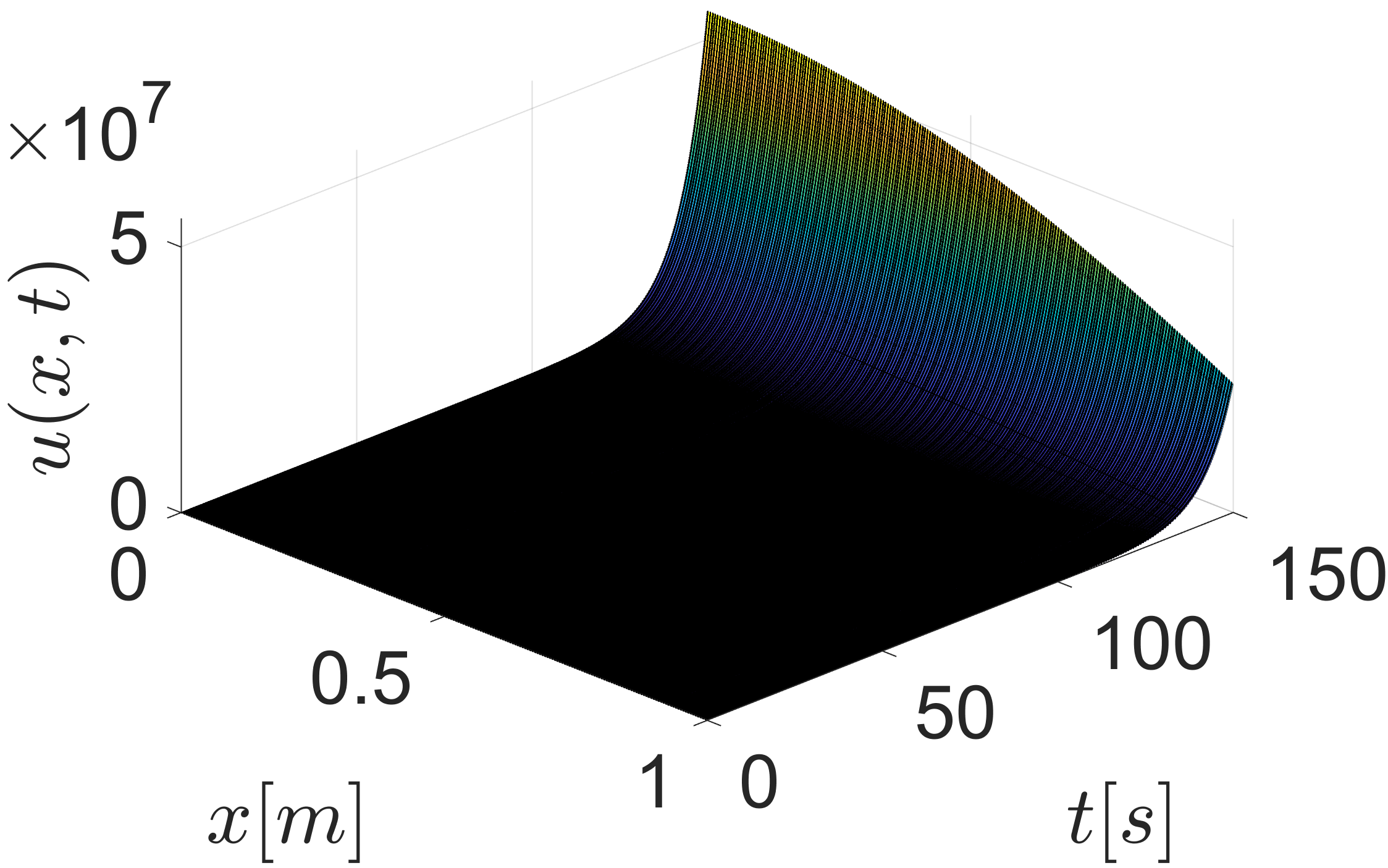}} \\\subfloat[]{\includegraphics[scale=.11]{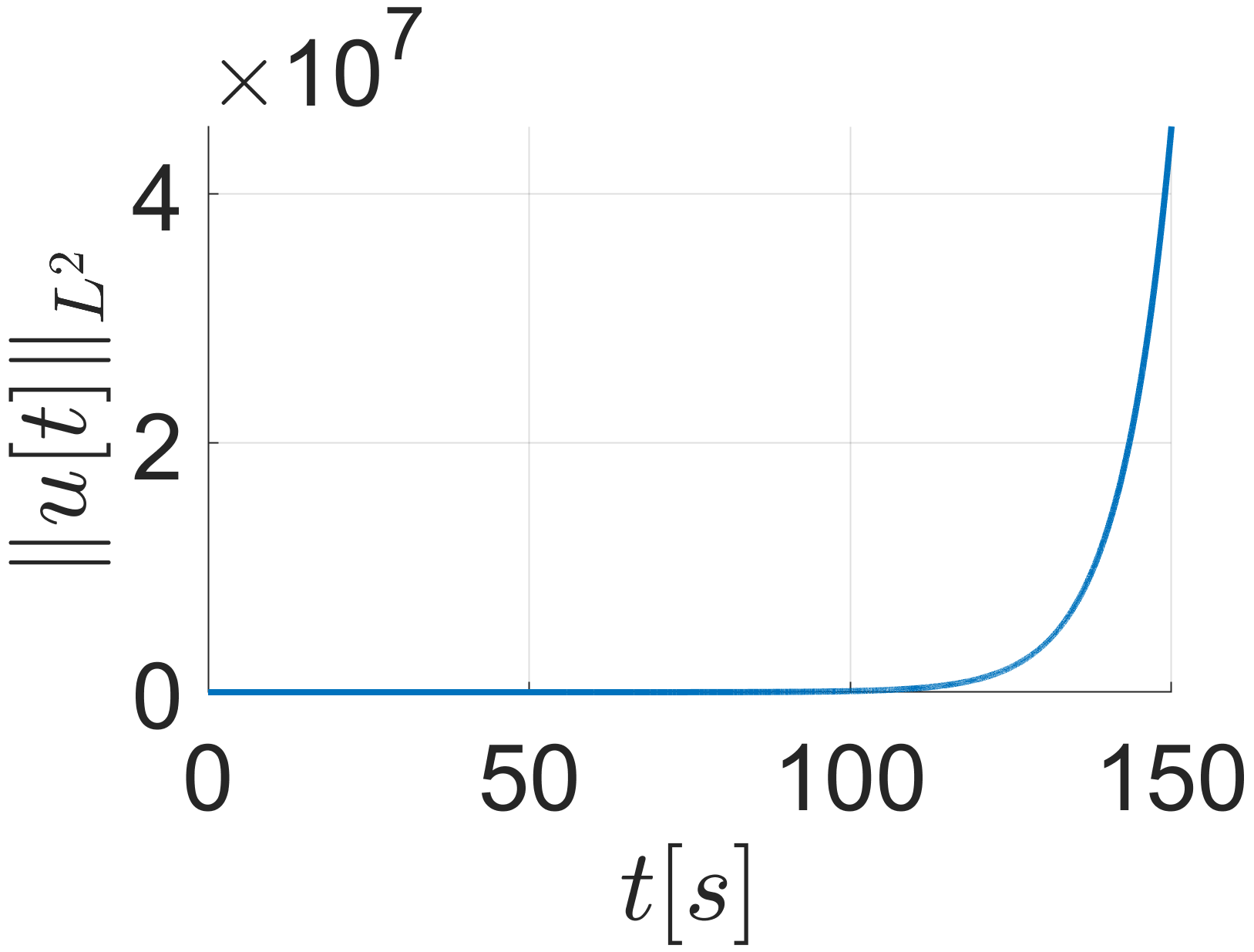} }
\caption{Results for open-loop plant with $\varepsilon=0.1,\lambda=0.25,q=2.3$ and $u[0]=10x^2(x-1)^{2}$.  (a) $u(x,t).$ (b) $\Vert u[t]\Vert.$}
\end{figure}

\begin{figure}
\centering
\subfloat[]{\includegraphics[scale=.11]{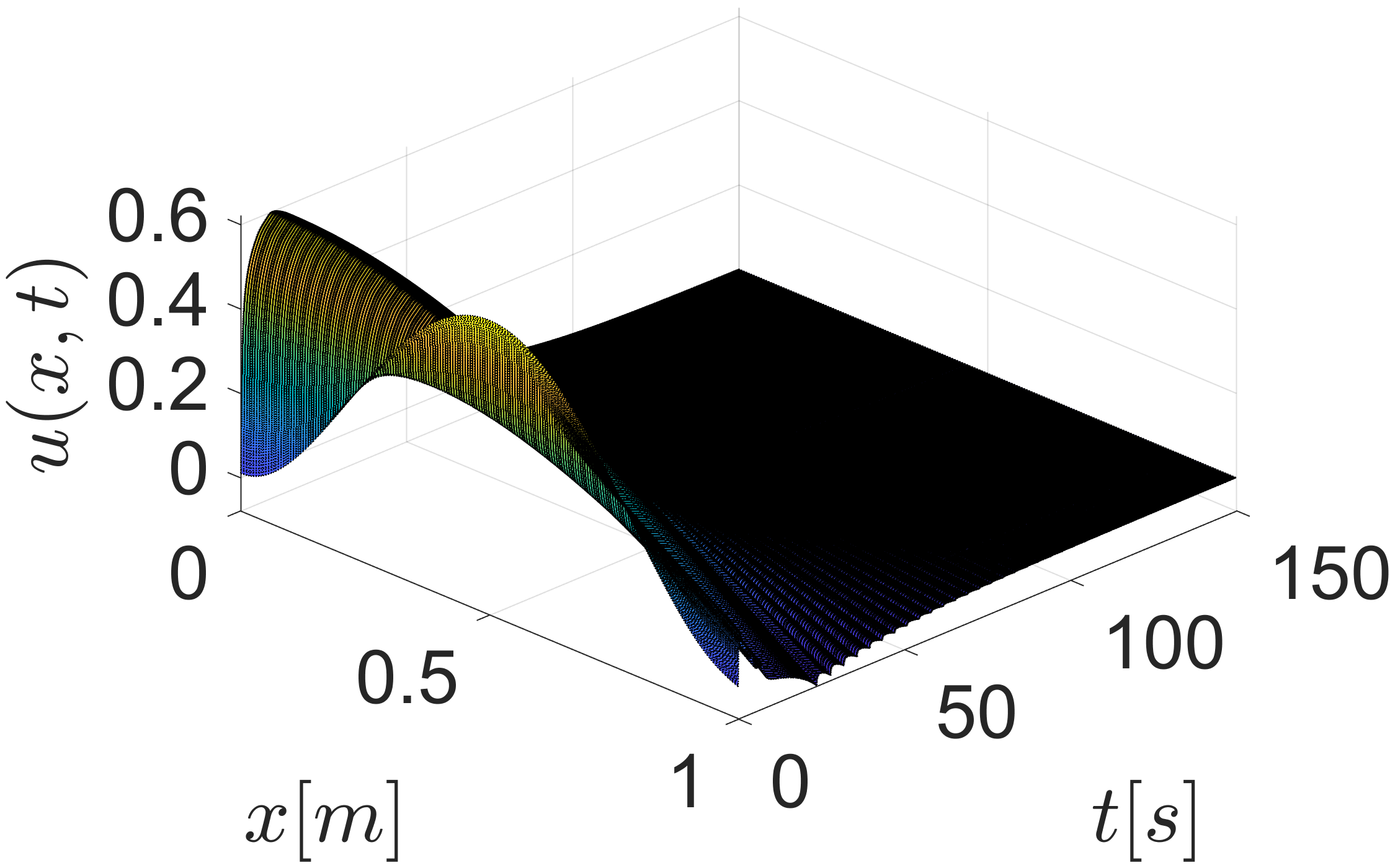}} \\\subfloat[]{\includegraphics[scale=.11]{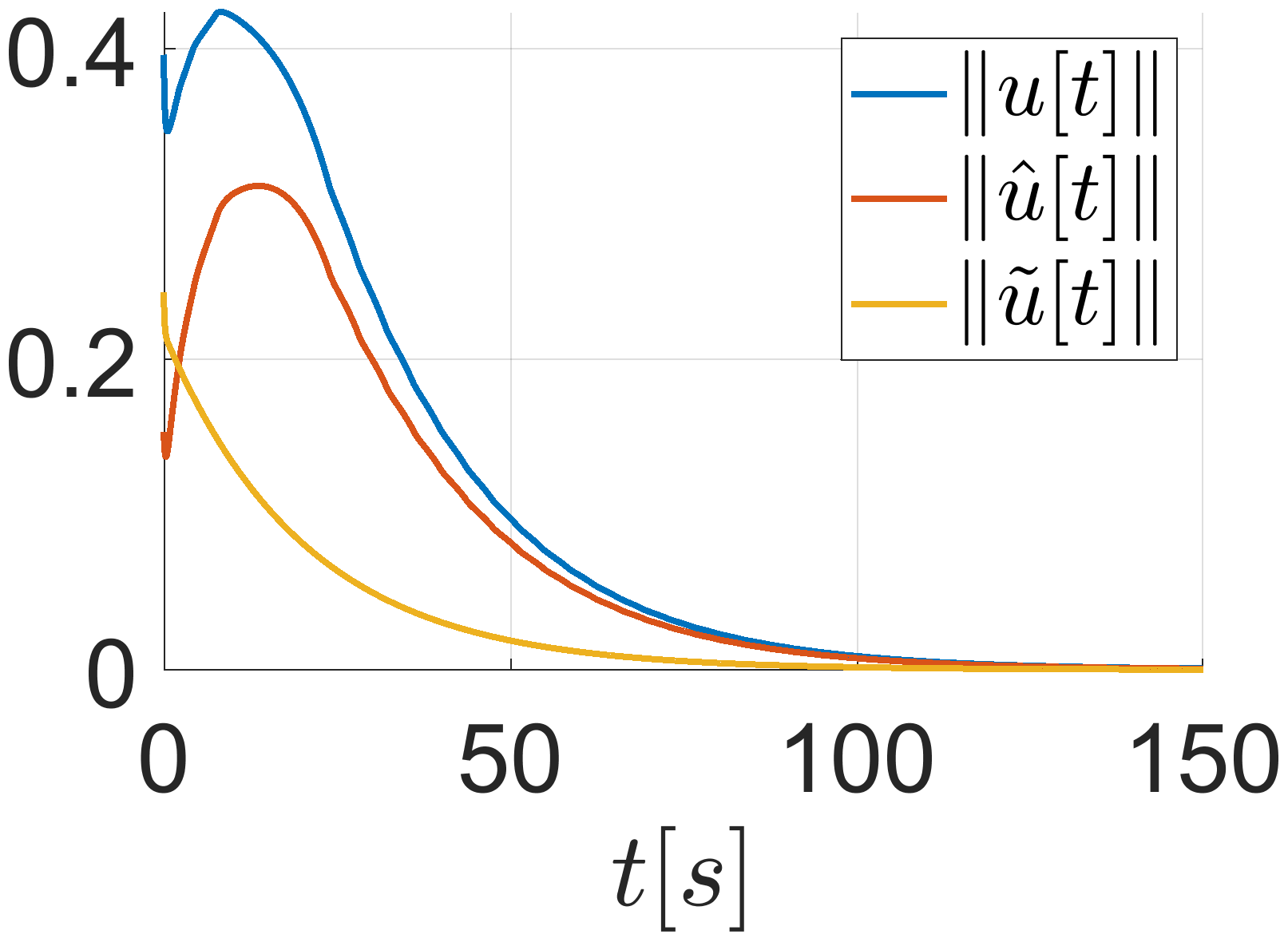}}
\caption{Results for the event-triggered closed-loop system with $\varepsilon=0.1,\lambda=0.25,q=2.3,m(0)=-10^{-4},\eta=1$, $u[0]=10x^2(x-1)^{2}$ and $\hat{u}[0]=5x^2(x-1)^2+5x^3(x-1)^3$ (a) $u(x,t)$. (b) $\Vert u[t]\Vert,
\Vert\hat{u}[t]\Vert\text{ and }\Vert\tilde{u}[t]\Vert.$ }
\end{figure}

\begin{figure}
\centering
\subfloat[]{\includegraphics[scale=.095]{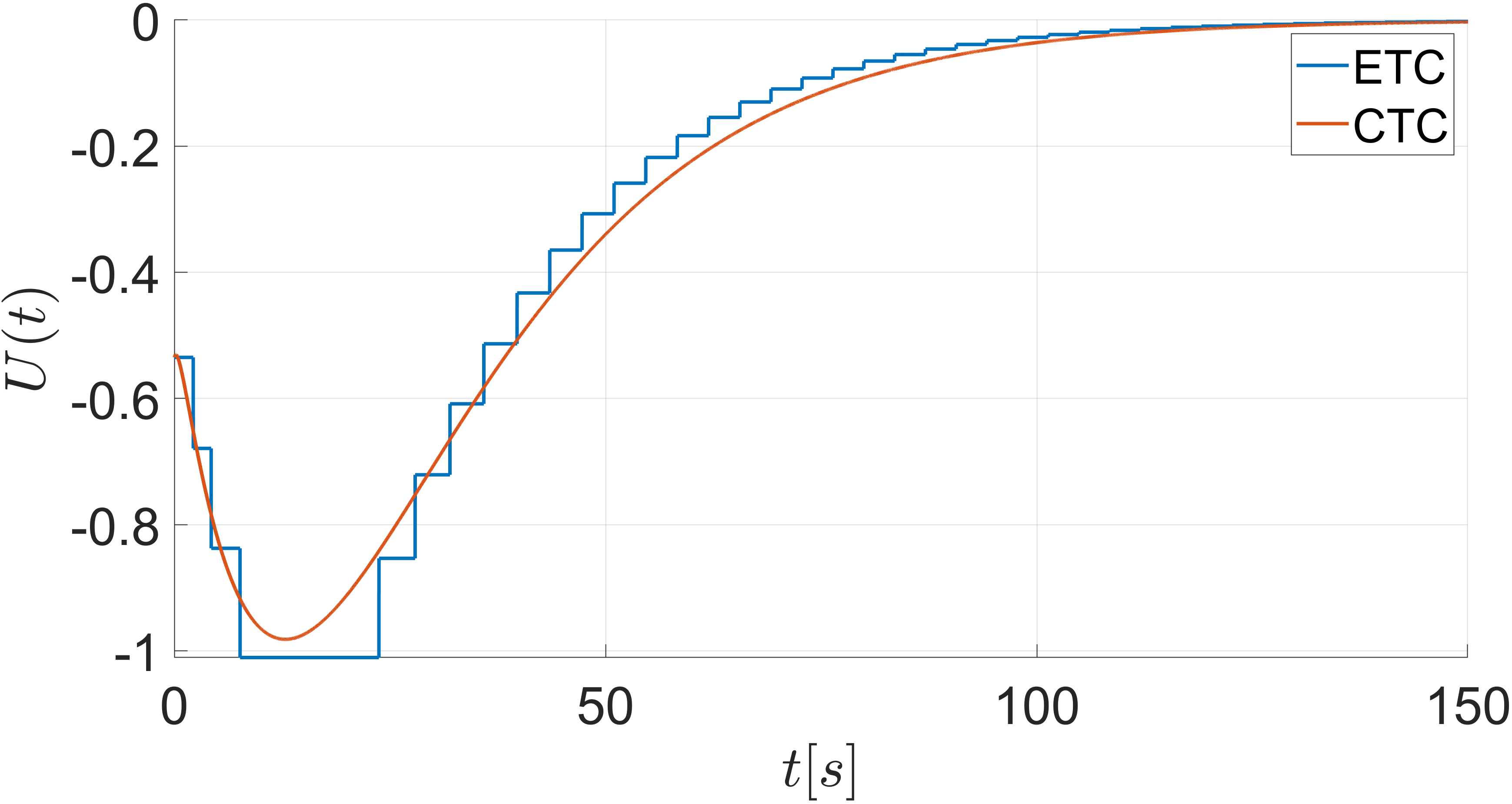}}
\hspace{5pt}\subfloat[]{\includegraphics[scale=.45]{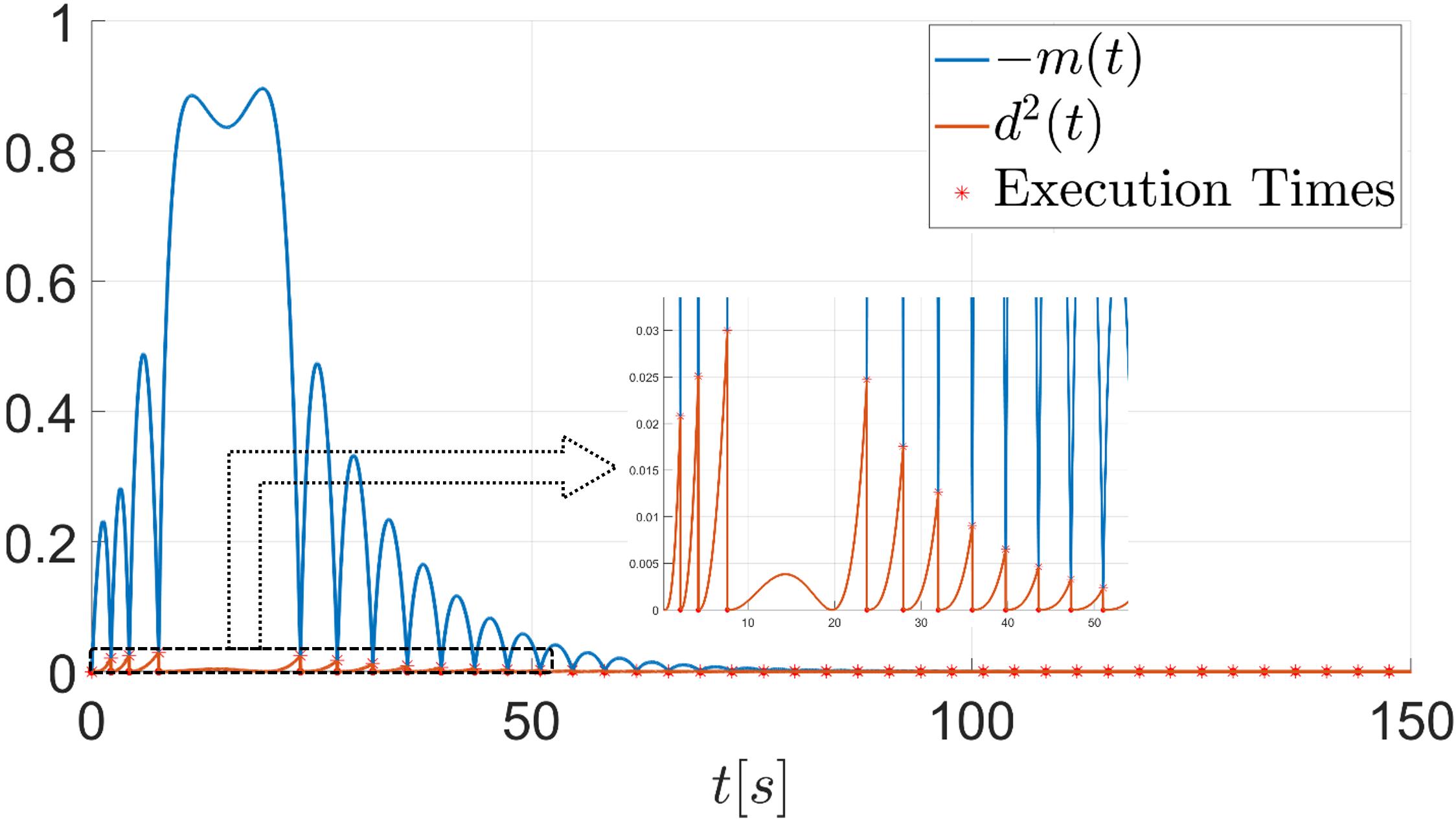}}
\caption{(a) The event-triggered control (ETC) input for the system considered in Fig. 3 along with the corresponding continuous-time control (CTC) input. (b) Trajectories involved in the triggering condition \eqref{obetbc1} for the system in Fig. 3.}
\end{figure}

\begin{figure}
\centering
\includegraphics[scale=.1]{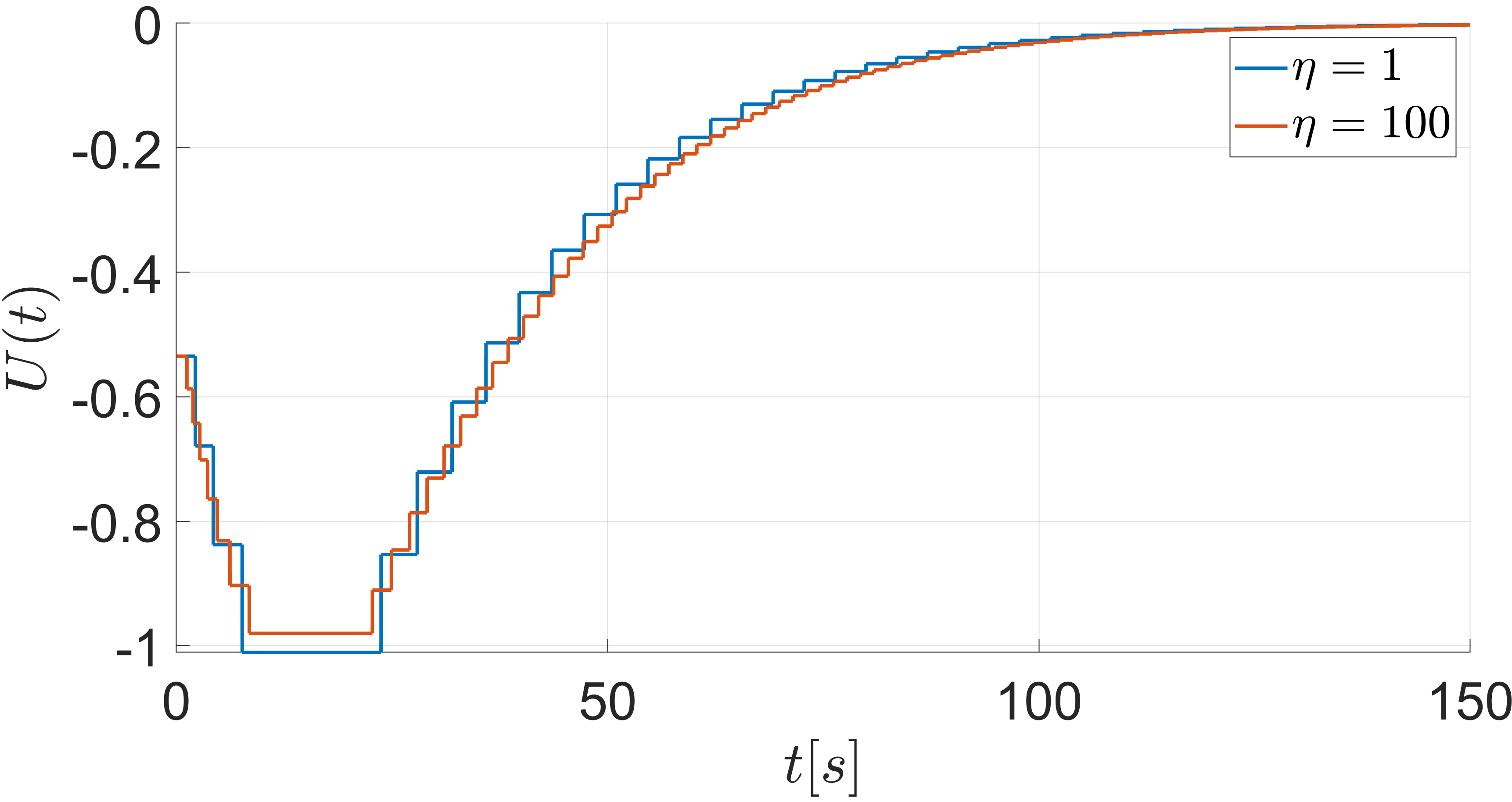}
\caption{Comparison of ETC input $U(t)$ for different $\eta$: $\eta=1\text{ and }\eta=100$, for the same system considered in Fig. 3.}
\end{figure}

\begin{figure}
\centering
\includegraphics[scale=.1]{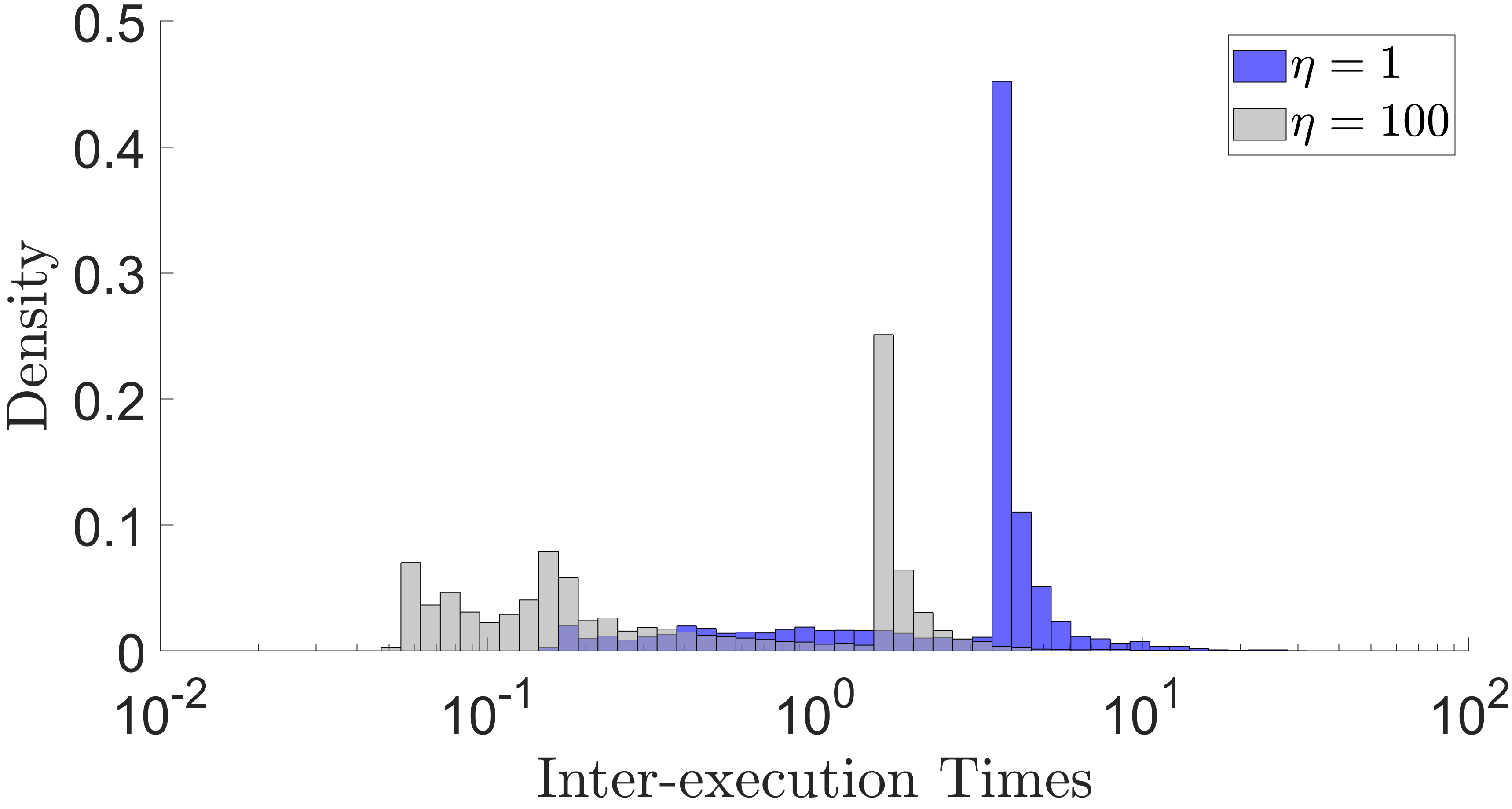}
\caption{Density of the inter-execution times (logarithmic scale) computed for 100 different initial conditions: $u[0]=x^2(x-1)^2\sin(n\pi x), n=1,\ldots,100$ and $\hat{u}[0]=2u[t]$ on a time frame of $150\text{ }s$, for the same system considered in Fig. 3 with $\eta=1$ and $\eta=100$.}
\end{figure}

\section{Conclusion}
This paper has proposed an event-triggered output feedback boundary control strategy for a class of reaction-diffusion systems with Robin boundary actuation. We have used a dynamic event triggering condition to determine when the control value needs to be updated. Under the proposed strategy, we have proved the existence of a minimal-dwell time between two updates independent of initial conditions, which excludes Zeno behavior. Further, we have shown the well-posedness of the closed-loop system  and its global $L^{2}$-exponential convergence to the equilibrium point.

We may consider periodic event-triggered boundary control (PETC) of reaction-diffusion systems under full-state and output feedback settings in future work. The idea is to evaluate the triggering condition periodically and to decide, at every sampling instant, whether the feedback loop needs to be closed. PETC is highly desirable as it not only guarantees a minimal dwell-time equal to the sampling period but also provides a more realistic approach toward digital implementations while reducing the utilization of computational resources.

\section*{Appendix A \\Proof of Proposition \ref{prop1}}

 Considering Remark \ref{rem1}, it can be deduced that the solution of  target system \eqref{ctots} satisfies  $\Vert \tilde{w}[t]\Vert\leq e^{-\varepsilon \mu^{2}t}\Vert\tilde{w}[0]\Vert$, which implies the global exponential stability in $L^{2}$-sense.

Let us show the gain kernel $P(x,y)$ and the observer gains $p_{1}(x)$  and $p_{10}$, which transform \eqref{ctoe} into \eqref{ctots} via \eqref{ctobt}, are given by \eqref{solP} and \eqref{p1}, respectively. This proves Proposition \ref{prop1}. 

Taking the time derivative of  \eqref{ctobt} along the solutions of \eqref{ctoe} and applying integration by parts twice, we can show that
\begin{equation}\label{ctobtr1}
\begin{split}
\tilde{u}_{t}(x,t)=&\tilde{w}_{t}(x,t)-\varepsilon P(x,x)\tilde{w}_{x}(x,t)\\&+\varepsilon P(x,0)\tilde{w}_{x}(0,t)+\varepsilon P_{y}(x,x)\tilde{w}(x,t)\\&-\varepsilon P_{y}(x,0)\tilde{w}(0,t)-\varepsilon\int_{0}^{x}P_{yy}(x,y)\tilde{w}(y,t)dy.
\end{split}
\end{equation}Differentiating \eqref{ctobt} w.r.t $x$ and using Leibnitz differentiation rule, we can obtain that
\begin{equation}\label{ctobtr2}
\begin{split}
\tilde{u}_{x}(x,t)=\tilde{w}_{x}(x,t)-P(x,x)\tilde{w}(x,t)-\int_{0}^{x}P_{x}(x,y)\tilde{w}(y,t)dy,
\end{split}
\end{equation}\begin{equation}\label{ctobtr3}
\begin{split}
\tilde{u}_{xx}(x,t)=&\tilde{w}_{xx}(x,t)-\frac{dP(x,x)}{dx}\tilde{w}(x,t)-P(x,x)\tilde{w}_{x}(x,t)\\&-P_{x}(x,x)\tilde{w}(x,t)-\int_{0}^{x}P_{xx}(x,y)\tilde{w}(y,t)dy.
\end{split}
\end{equation}
Therefore, from \eqref{ctoee1},\eqref{ctobt},\eqref{ctobtr1} and \eqref{ctobtr3}, we can show that
\begin{equation}\label{ap1}
\begin{split}
&0=\big(p_{1}(x)-\varepsilon P_{y}(x,0)\big)\tilde{w}(0,t)-\bigg(\lambda-2\varepsilon\frac{dP(x,x)}{dx}\bigg)\tilde{w}(x,t)\\&+\int_{0}^{x}\bigg(\varepsilon P_{xx}(x,y)-\varepsilon P_{yy}(x,y)+\lambda P(x,y)\bigg)\tilde{w}(y,t)dy.
\end{split}
\end{equation}Let us choose $P_{xx}(x,y)-P_{yy}(x,y)=-\frac{\lambda}{\varepsilon}P(x,y),\frac{dP(x,x)}{dx}=\frac{\lambda}{2\varepsilon},\text{ and }p_{1}(x)=\varepsilon P_{y}(x,0)$
so that \eqref{ap1} is valid for any $\tilde{w}$. Further, let us choose $p_{10}=P(0,0)$ so that the boundary conditions \eqref{ctoee2} and \eqref{ctotse2} are satisfied, and choose $P(1,1)=0$ and $P_{x}(1,y)=-qP(1,y)$  so that the conditions \eqref{ctoee3} and \eqref{ctotse3} are met. Therefore, the gain kernel $P(x,y)$ in \eqref{ctobt} as a whole should satisfy the following PDE:  
\begin{subequations}\label{ctok}
\begin{align}\label{ctoke1}
P_{xx}(x,y)-P_{yy}(x,y)&=-\frac{\lambda}{\varepsilon}P(x,y),
\\
P_{x}(1,y)&=-qP(1,y),
\\
P(x,x)&=\frac{\lambda}{2\varepsilon}(x-1).
\end{align}
\end{subequations}
It can be shown that the change of variables $x=1-\bar{y}$ and $y=1-\bar{x}$ on \eqref{ctok} leads to the same PDE (108-110) in \cite{smyshlyaev2004closed} to which the explicit solution has been obtained. Therefore, the solution to \eqref{ctok} can be shown to be given by \eqref{solP}. Above we have obtained $p_{1}(x)=\varepsilon P_{y}(x,0)$ and $p_{10}=P(0,0)$, which are the same as stated in Proposition \ref{prop1}.$\qed$

\section*{Appendix B \\Proof of Proposition \ref{prop2}}
 Let us show that the gain kernel $K(x,y)$  and the control law $U(t)$, which transform \eqref{cto} into \eqref{etots}-\eqref{rt} via \eqref{ctbt}, are given by \eqref{ctcks} and \eqref{ctcl}, respectively. This proves Proposition \ref{prop2}.   
 
Taking the time derivative of  \eqref{ctbt} along the solutions of \eqref{cto} and applying integration by parts twice, we can show that\begin{equation}\label{etobtr1}
\begin{split}
&\hat{w}_{t}(x,t)=\hat{u}_{t}(x,t)-\lambda\int_{0}^{x}K(x,y)\hat{u}(y,t)dy
\\&-\int_{0}^{x}K(x,y)p_{1}(y)dy\tilde{w}(0,t)
-\varepsilon K(x,x)\hat{u}_{x}(x,t)\\&+\varepsilon K(x,0)\hat{u}_{x}(0,t)+\varepsilon K_{y}(x,x)\hat{u}(x,t)\\&-\varepsilon K_{y}(x,0)\hat{u}(0,t)-\varepsilon \int_{0}^{x}K_{yy}(x,y)\hat{u}(y,t)dy,
\end{split}
\end{equation}Differentiating \eqref{ctbt} w.r.t $x$ and using Leibnitz differentiation rule, we can obtain that\begin{equation}\label{etobtr2}
\begin{split}
\hat{w}_{x}(x,t)=\hat{u}_{x}(x,t)-K(x,x)\hat{u}(x,t)-\int_{0}^{x}K_{x}(x,y)\hat{u}(y,t)dy,
\end{split}
\end{equation}\begin{equation}\label{etobtr3}
\begin{split}
\hat{w}_{xx}(x,t)=&\hat{u}_{xx}(x,t)-\frac{dK(x,x)}{dx}\hat{u}(x,t)-K(x,x)\hat{u}_{x}(x,t)\\&-K_{x}(x,x)\hat{u}(x,t)-\int_{0}^{x}K_{xx}(x,y)\hat{u}(y,t)dy.
\end{split}
\end{equation}
Therefore, from \eqref{p1},\eqref{ctbt},\eqref{etotse1},\eqref{gt},\eqref{etobtr1} and \eqref{etobtr3}, we can show
\begin{equation}\label{rnd}
\begin{split}
&0=\Big(\lambda+2\varepsilon\frac{dK(x,x)}{dx}\Big)\hat{u}(x,t)dy-\varepsilon K_{y}(x,0)\hat{u}(0,t)\\&+\int_{0}^{x}\bigg(\varepsilon K_{xx}(x,y)-\varepsilon K_{yy}(x,y)-\lambda K(x,y)\bigg)\hat{u}(y,t)dy.
\end{split}
\end{equation}Let us choose $K_{xx}(x,y)- K_{yy}(x,y)=\frac{\lambda}{\varepsilon}K(x,y),K_{y}(x,0)=0,\text{ and, }\frac{dK(x,x)}{dx}=-\frac{\lambda}{2\varepsilon}$ so that \eqref{rnd} holds for any $\hat{u}$.  Further, let us choose $K(0,0)=0,$ so that the boundary conditions \eqref{ctoe2} and \eqref{etotse2} are met, and choose $U(t)=\int_{0}^{1}\Big(rK(1,y)+K_{x}(1,y)\Big)\hat{u}(y,t)dy$ so that the boundary conditions \eqref{ctoe3} and \eqref{etotse3} are satisfied. Therefore, the gain kernel $K(x,y)$ in \eqref{ctbt} as a whole should satisfy the following PDE:
\begin{subequations}\label{ctck}
\begin{align}\label{ctcke1}
K_{xx}(x,y)- K_{yy}(x,y)&=\frac{\lambda}{\varepsilon}K(x,y),
\\
K_{y}(x,0)&=0,
\\
K(x,x)&=-\frac{\lambda}{2\varepsilon}x.
\end{align}
\end{subequations}
The solution to \eqref{ctck} is given by \eqref{ctcks}\cite{krstic2008boundary}. Further, the control law obtained above is the same as \eqref{ctcl}.$\qed$

\section*{Appendix C \\Proof of Proposition \ref{prop3}}

Subject to Assumption \ref{ass1}, let us choose parameters $\delta_{1},\delta_{2}>0$ such that\begin{equation}\label{pps1}
\varepsilon\min\Big\{r-\frac{1}{2},\frac{1}{2}\Big\}-\frac{5\lambda}{8\delta_{1}}-\frac{\Vert g\Vert^{2}}{\delta_{2}}\geq 0,
\end{equation}and $H>0$ such that\begin{equation}\label{pps2}
H\varepsilon\min\Big\{q-\frac{1}{2},\frac{1}{2}\Big\}-\frac{5\lambda\delta_{1}}{8}-\frac{5\delta_{2}}{4}\geq 0.
\end{equation}Here $g(x)$ and $r$ are given by \eqref{gt} and \eqref{rt}, respectively. Note that $r>1/2$ due to Assumption \ref{ass1}. Then, let us consider the following Lyapunov candidate
\begin{equation}
\mathcal{V}=\frac{H}{2}\int_{0}^{1}\tilde{w}^{2}(x,t)dx+\frac{1}{2}\int_{0}^{1}\hat{w}^{2}(x,t)dx,
\end{equation}
where $\tilde{w}$ and $\hat{w}$ are the systems described by \eqref{ctots} and\eqref{etots}, respectively. We can show that
\begin{equation}\label{pdlf}
\begin{split}
&\dot{\mathcal{V}}=-H\varepsilon q\tilde{w}^{2}(1,t)-H\varepsilon\int_{0}^{1}\tilde{w}^{2}_{x}(x,t)dx\\&-r\varepsilon \hat{w}^{2}(1,t)+\frac{\lambda}{2}\hat{w}(0,t)\tilde{w}(0,t)\\&-\varepsilon \int_{0}^{1}\hat{w}_{x}^{2}(x,t)dx+\int_{0}^{1}g(x)\hat{w}(x,t)dx\tilde{w}(0,t).
\end{split}
\end{equation}
From Young's and Cauchy-Schwarz inequalities, we can obtain that
\begin{equation}\label{pf1}
\frac{\lambda}{2}\hat{w}(0,t)\tilde{w}(0,t)\leq\frac{\lambda}{4\delta_{1}}\hat{w}^{2}(0,t)+\frac{\lambda\delta_{1}}{4}\tilde{w}^{2}(0,t),
\end{equation}
\begin{equation}\label{pf2}
\begin{split}
\int_{0}^{1}g(x)\hat{w}(x,t)dx\tilde{w}(0,t)\leq&\frac{\Vert g\Vert^{2}}{2\delta_{2}}\Vert \hat{w}[t]\Vert^{2}+\frac{\delta_{2}}{2}\tilde{w}^{2}(0,t).
\end{split}
\end{equation}
Therefore, using \eqref{pf1} and \eqref{pf2}, we can write \eqref{pdlf} as
\begin{equation}\label{plfe1}
\begin{split}
\dot{\mathcal{V}}\leq&-H\varepsilon q\tilde{w}^{2}(1,t)-H\varepsilon\Vert\tilde{w}_{x}[t]\Vert^{2}-r\varepsilon \hat{w}^{2}(1,t)\\&+\frac{\lambda }{4\delta_{1}}\hat{w}^{2}(0,t)+\frac{\lambda\delta_{1}}{4}\tilde{w}^{2}(0,t)\\&
-\varepsilon \Vert\hat{w}_{x}[t]\Vert^{2}+\frac{\Vert g\Vert^{2}}{2\delta_{2}}\Vert\hat{w}[t]\Vert^{2}+\frac{\delta_{2}}{2}\tilde{w}^{2}(0,t).
\end{split}
\end{equation}
From Agmon's and Young's inequalities, we have that
\begin{equation}
\tilde{w}^{2}(0,t)\leq \tilde{w}^{2}(1,t)+\Vert \tilde{w}[t]\Vert^{2}+\Vert\tilde{w}_{x}[t]\Vert^{2},
\end{equation}\begin{equation}
\hat{w}^{2}(0,t)\leq \hat{w}^{2}(1,t)+\Vert \hat{w}[t]\Vert^{2}+\Vert\hat{w}_{x}[t]\Vert^{2}.
\end{equation}
Therefore, we can show using \eqref{plfe1} that
\begin{equation}\label{ac1}
\begin{split}
\dot{\mathcal{V}}\leq& -\Big(H\varepsilon q-\frac{\lambda\delta_{1}}{4}-\frac{\delta_{2}}{2}\Big)\tilde{w}^{2}(1,t)\\&-\Big(H\varepsilon-\frac{\lambda\delta_{1}}{4}-\frac{\delta_{2}}{2}\Big)\Vert\tilde{w}_{x}[t]\Vert^{2}\\&+\Big(\frac{\lambda\delta_{1}}{4}+\frac{\delta_{2}}{2}\Big)\Vert\tilde{w}[t]\Vert^{2}
-\Big(r\varepsilon-\frac{\lambda}{4\delta_{1}}\Big)\hat{w}^{2}(1,t)\\&-\Big(\varepsilon -\frac{\lambda }{4\delta_{1}}\Big)\Vert\hat{w}_{x}[t]\Vert^{2}+\Big(\frac{\lambda}{4\delta_{1}}+\frac{\Vert g\Vert^{2}}{2\delta_{2}}\Big)\Vert\hat{w}[t]\Vert^{2}.
\end{split}
\end{equation}From Poincar\'e Inequality, we have that
\begin{equation}\label{ac2}
-\Vert \tilde{w}_{x}[t]\Vert^{2}\leq \frac{1}{2}\tilde{w}^{2}(1,t)-\frac{1}{4}\Vert \tilde{w}[t]\Vert^{2},
\end{equation}\begin{equation}\label{ac3}
-\Vert \hat{w}_{x}[t]\Vert^{2}\leq \frac{1}{2}\hat{w}^{2}(1,t)-\frac{1}{4}\Vert \hat{w}[t]\Vert^{2}.
\end{equation}
Furthermore, we have from \eqref{pps1}  and \eqref{pps2} that

\begin{equation}\label{ac4}
H\varepsilon-\frac{\lambda\delta_{1}}{4}-\frac{\delta_{2}}{2}>0\text{ and }\varepsilon -\frac{\lambda }{4\delta_{1}}>0.
\end{equation} 
Therefore, using \eqref{ac1}-\eqref{ac4}, we can show that
\begin{equation}\label{imc}
\begin{split}
\dot{\mathcal{V}}\leq &-\Big(H\varepsilon(q-\frac{1}{2})-\frac{\lambda\delta_{1}}{8}-\frac{\delta_{2}}{4}\Big)\tilde{w}^{2}(1,t)\\&-\Big(\frac{H\varepsilon}{4}-\frac{5\lambda\delta_{1}}{16}-\frac{5\delta_{2}}{8}\Big)\Vert\tilde{w}[t]\Vert^{2}\\&
-\Big(\varepsilon (r-\frac{1}{2})-\frac{\lambda }{8\delta_{1}}\Big)\hat{w}^{2}(1,t)\\&-\Big(\frac{\varepsilon}{4}-\frac{5\lambda }{16\delta_{1}}-\frac{\Vert g\Vert^{2}}{2\delta_{2}}\Big)\Vert\hat{w}[t]\Vert^{2}.
\end{split}
\end{equation}
From \eqref{pps1}  and \eqref{pps2}, we have that

\begin{equation}
H\varepsilon(q-\frac{1}{2})-\frac{\lambda\delta_{1}}{8}-\frac{\delta_{2}}{4}>0 \text{ and }\varepsilon (r-\frac{1}{2})-\frac{\lambda }{8\delta_{1}}>0,
\end{equation}Thus, it follows from \eqref{imc} that 
\begin{equation}\label{tie}
\begin{split}
\dot{\mathcal{V}}\leq&-\Big(\frac{H\varepsilon}{4}-\frac{5\lambda\delta_{1}}{16}-\frac{5\delta_{2}}{8}\Big)\Vert\tilde{w}[t]\Vert^{2}\\&
-\Big(\frac{\varepsilon}{4}-\frac{5\lambda }{16\delta_{1}}-\frac{\Vert g\Vert^{2}}{2\delta_{2}}\Big)\Vert\hat{w}[t]\Vert^{2}.
\end{split}
\end{equation}
Again, we have from \eqref{pps1}  and \eqref{pps2} that

\begin{equation}
\vartheta_{1}=\frac{H\varepsilon}{4}-\frac{5\lambda\delta_{1}}{16}-\frac{5\delta_{2}}{8}>0,
\end{equation}
\begin{equation}
\vartheta_{2}=\frac{\varepsilon}{4}-\frac{5\lambda }{16\delta_{1}}-\frac{\Vert g\Vert^{2}}{2\delta_{2}}>0.
\end{equation}Therefore, \eqref{tie} can be written as

\begin{equation}\label{lstw}
\dot{\mathcal{V}}\leq-\frac{\min\big\{\vartheta_{1},\vartheta_{2}\big\}}{\max\big\{H/2,1/2\big\}}\mathcal{V}.
\end{equation}Hence, from standard arguments, we can state that the closed-loop system which consists of the plant \eqref{ctp} and the observer \eqref{cto} with the continuous-time control law \eqref{ctcl}, is globally exponentially stable in $L^{2}$-sense.$\qed$
\bibliographystyle{IEEEtran}
\bibliography{main}
\end{document}